\documentclass[preprint, 1p, authoryear]{elsarticle}
\title{On pore--scale modeling and simulation of reactive transport in 3D geometries}

\usepackage{fullpage,graphicx,amsmath,amssymb, color, multirow}
\usepackage{empheq}
\usepackage{natbib}
\usepackage{tikz}
\usetikzlibrary{shapes,arrows}
\usepackage{color}

\usepackage{caption}
\usepackage[labelformat=simple]{subcaption}

\journal{}
\begin{document}

\begin{frontmatter}

\author[Fraunhofer,Saudi]{Oleg Iliev\corref{cor1}}
\ead{iliev@itwm.fraunhofer.de}

\author[Fraunhofer,DHI]{Zahra Lakdawala}
\ead{zla@dhigroup.com}

\author[Fraunhofer]{Katherine~H.~L.~Leonard}
\ead{leonard@itwm.fraunhofer.de}

\author[Bulgaria]{Yavor Vutov}
\ead{yavor@parallel.bas.bg}

\cortext[cor1]{Corresponding author}
\address[Fraunhofer]{Fraunhofer Institute for Industrial Mathematics ITWM, Germany.}
\address[Saudi]{Numerical Porous Media SRI Center, King Abdullah University of Science and Technology, Kingdom of Saudi Arabia.}
\address[DHI]{DHI-WASY GmbH, Berlin, Germany.}
\address[Bulgaria]{Institute of Information and Communication Technologies, Department of Scientific Computations, Bulgarian Academy of Sciences, Sofia, Bulgaria.}

\begin{abstract}
Pore--scale modeling and simulation of reactive flow in porous media  has a range of diverse applications, and  poses a number of research challenges. 
It is known that the morphology of a  porous medium has significant influence on the local flow rate, which can have a substantial impact on the rate of chemical reactions. While  there are a large number of papers and software tools dedicated  to simulating   either fluid flow in 3D computerized tomography (CT)  images or reactive flow using pore--network models, little attention to date has been focused on the pore--scale simulation of sorptive transport in 3D CT images, which is the specific focus of this paper. 
Here we first present an algorithm for the simulation of such reactive flows directly on images, which is implemented in a sophisticated software package.  We then use this software to present numerical results in two  resolved geometries, illustrating the importance of pore--scale simulation and the flexibility of our software package. 
\end{abstract}

\end{frontmatter}

\section{Introduction}

Understanding and controlling reactive flow in porous media is important for a number of environmental and industrial applications, including oil recovery, fluid filtration and purification, combustion and hydrology.  Traditionally, the majority  of  theoretical  and experimental research into transport within porous media has been carried out at macroscopic (Darcy) scale.  
Despite the progress in developing devices to perform experimental measurements at the pore--scale, experimental characterization of the pore--scale velocity, pressure and solute fields is still a challenging task. Due the influence of the pore--scale geometry on the processes of interest, direct numerical simulation (DNS) promises to be a very useful computational tool in  a wide range of fields, and in combination with experimental studies, can be used to determine quantities of interest that are not experimentally quantifiable \citep{Blunt2013Porescale}.

Significant progress  over the past 10~--~15 years  in the pore--scale simulation of single phase flow has resulted in the  computation of  permeability tensors for natural and technical porous media becoming  a standard procedure. A number of academic as well as commercial software tools, capable of processing 3D CT images in addition to virtually generating porous media, are available,  for example Aviso, GeoDict and Ingrain.  Most of those software tools have the additional ability of simulating  two--phase immiscible flow at the pore--scale directly on  a computational domain obtained through the segmentation of   3D CT images, often  using  the lattice Boltzmann (LB), the level set  or  volume of fluid methods.  In contrast, substantially less work on the DNS of reactive flow has been performed, and only a few  software tools with this capability exist. 
A limited number of computational studies examining  reactive transport where the reactions only occur within the fluid phase (and not at a surface) exist \citep{Molins2012Investigation, Shen2011High}.    
In contrast, the literature and computational tools examining full 3D reactive flow where  the reactions occur at the pore wall (surface reactions), is sparse. The majority of existing  studies and available numerical simulation packages describing pore--scale sorptive transport are based on pore--network mathematical models (see, for example,  \citet{Raoof2011Upscaling, Adler2013} and literature therein), where the geometry needs to be converted into an idealized series of connected pores and throats to represent the porous medium. However, during this process, information on the morphology of the underlying media can be lost (see, for example, \citet{Raoof2011Upscaling} and \citet{Lichtner2007Upscaling}). In this paper we present an algorithm and a sophisticated software package, called Pore--Chem, which 
uses cell centered finite volume (FV) methods to numerically solve 3D solute transport with surface 
reactions at the pore--scale. 
In particular the software package has the ability to solve the systems of equations modeling colloidal 
reactive transport on a geometrical domain obtained directly through imaging techniques, such as 
computerized tomography, which allows for a very accurate spatial description of the computational 
domain.

In this paper we describe the transport of a generic solute through the  Navier--Stokes (NS) system of equations coupled to  a convection--diffusion  (CD) equation.  The CD equation is complemented by boundary conditions which describe various types of surface reactions comprising a  Robin  boundary condition for the solute coupled to an ordinary differential equation (ODE) describing the dynamics of the adsorption at the pore wall.  
To represent the computational domain,  a voxelized geometry is considered, where each individual voxel is either solid or fluid. 
 The solute transport model is chosen due to its applicability to a broad range of problems.
In addition, our  goal is to describe the transport and reaction of sub--micron particles, for which inertial effects of the individual particles are negligible. 
Discrete models, where  each  particle is modeled as an individual entity,  are necessary when  considering larger particles, for example with a radius greater  than one micron,  for which inertial effects become important.  Several commercial software packages, for example GeoDict \citep{Geodict}, solve a range of discrete mathematical models describing colloid transport and adsorption. However, numerical simulation of these models is  often significantly more computationally expensive  than a continuous mathematical model and accounting for different reaction kinetics is usually not possible in such packages.

Reactive flow in porous media is intrinsically a multiscale problem. The goal of our developments is to support    problems where scale separation is   possible and in cases where it is not possible. The first case, where the  separation of scales is viable,  is usually the focus  of asymptotic homogenization theory. In the second case, when  scale separation is not possible,  numerical upscaling methods like  multiscale finite element methods are often applied.  During the homogenization procedure, when applicable, certain assumptions are imposed, allowing for the derivation of  macroscopic (Darcy scale) equations from the microscale formulation, with effective parameters, such as the permeability and the effective reaction rate, obtained through solution of a  cell problem \citep{hornung1997homogenization}.  
  A number of studies have employed homogenization theory to derive a macroscopic description of sorptive reactive transport for particular parameter regimes.   The homogenization of solute transport in porous media in the presence of surface reactions has been performed for both high P\'eclet numbers   (convection dominated regime) \citep{Allaire2010Homogenization,Allaire2010Twoscale,Allaire2012}, and when the P\'eclet number  is of order one \citep{HornungMikelic,KumarPop,Knabner}.   
In addition to being able to solve cell problems in a number of settings, our software has the ability of solving a much broader class of problems at the pore--scale, without being restricted by the assumptions required during  homogenization.  Furthermore, it provides the possibility to study various different types of surface reactions  described by different kinetics.

The remainder of the paper is organized as follows. In Section~\ref{sec:Mathematical_Model} the 
mathematical model is presented and cast into dimensionless variables. The method of achieving a 
numerical solution to the system of equations is outlined in Section~\ref{sec:numerical_methods}, and 
illustrative results using this method are presented in Section~\ref{sec:Results}.  Finally, conclusions are drawn in 
Section~\ref{sec:Conclusion_Discussion}.

\section{Mathematical model}
\label{sec:Mathematical_Model}
We now detail  the mathematical model, which describes the transport and reaction of a generic solute at a 2D interface within a 3D pore--scale resolved geometry, where we assume that the number of solute particles is sufficiently large  that representation within a continuum framework is valid. 


Let us denote the spatial domain of interest by $\Omega$, an open subset of $\mathbb{R}^3$. We assume that we can decompose $\Omega$ into a solid domain, denoted by $\Omega_s$, and a fluid domain, denoted by $\Omega_f$, such that   $\Omega = \Omega_s \cup \Omega_f$. Denoting the  external boundary of our domain, being the closure of $\Omega$, by $\partial \Omega$, we partition this into an inlet, $\partial \Omega_{\text{in}}$, an outlet, $\partial \Omega_{\text{out}}$,  and  external walls, $\partial\Omega_{\text{wall}}$,  so that
\begin{equation}
\partial \Omega = \partial \Omega_{\text{in}} \cup \partial \Omega_{\text{out}} \cup \partial \Omega_{\text{wall}} .
\end{equation}
We note that, although we here consider only one inlet and one outlet, extension to consider multiple inlets and outlets 
 is straightforward. Finally, we denote the interfacial boundary between the fluid and solid portions of the domain by $\Gamma = \Omega_f \cap \Omega_s$. To allow for different types of reactive boundary conditions to be described, we decompose $\Gamma$ into $N \geq 1$  different boundary types as follows
\begin{equation}
\Gamma = \cup_{i=0}^{N-1} \Gamma_i ,
\end{equation}
where $\Gamma_i$ has distinct properties. Making such a decomposition enables us to attribute different reaction rates or kinetic descriptions to different portions of the domain, allowing for different types of solid material. 

In order to describe the flow of the water through the membrane, by appealing to the conservation of momentum,  the incompressible NS equations are used:
\begin{subequations}
\label{eq:Navier_Stokes_equations}
\begin{empheq}[right={\qquad \mathbf{x} \in \Omega_f, \: t > 0,}]{align}
\rho \left( \displaystyle\frac{\partial \mathbf{v} }{\partial t} + \mathbf{v} \cdot \nabla \mathbf{v} \right) &= - \nabla p + \mu\nabla \cdot \left( \nabla \mathbf{v}\right) ,\label{eq:Navier_Stokes_equations_eq1} \\
\nabla \cdot \mathbf{v} &= 0, \label{eq:Navier_Stokes_equations_eq2}
\end{empheq}
\end{subequations}
where $\mathbf{v}(\mathbf{x},t)$ and $p(\mathbf{x},t)$ are the velocity and pressure of the fluid respectively, while $\mu\geq0$ and $\rho \geq 0$ are the viscosity and the density of the fluid which we assume are constants \citep{bear1988dynamics,Acheson1995Elementary}. Suitable boundary conditions on $\partial \Omega$ are given by
\begin{subequations}
\label{eq:Navier_Stokes_boundary_conditions}
\begin{empheq}[right={\qquad t > 0.}]{align}
\mathbf{v} &= \textbf{V}_{\text{in}} , \qquad & \mathbf{x} &\in \partial \Omega_{\text{in}}, \\
p &=P_{\text{out}} 
\qquad & \mathbf{x}  &\in \partial \Omega_{\text{out}}, ~\\
\mathbf{v} &= \mathbf{0}, \qquad& \mathbf{x} &\in   \partial \Omega_{\text{wall}}, 
\end{empheq}
\end{subequations}
where $\textbf{V}_{\text{in}}$ is the inlet velocity with $\| \mathbf{V}_{\text{in}} \|>0$, $P_{\text{out}}$ is the pressure at the  outlet, and $\mathbf{n}$ is the outward pointing normal to the boundary $\partial \Omega$. Although here we have used no--slip and no--flux  flow conditions for the external walls,  symmetry or periodic boundary conditions can  also be imposed which may be more appropriate depending on the problem to be solved. 
 Further boundary conditions are required to be specified on the reminder of the boundary to $\Omega_f$, being the fluid--solid interface. To allow for the slip of flow along the fluid--solid interface, and the  inclusion of additional effects such as charged fluids or matrices,  we use the Navier--Maxwell slip conditions, given by
\begin{equation}
\mathbf{v} \cdot \mathbf{n} = 0, \qquad 
\mathbf{v} \cdot \mathbf{t} = \beta_i \mathbf{n} \cdot \left(\nabla \mathbf{v} + \left( \nabla \mathbf{v} \right)^{\mathrm{T}}\right) \cdot \mathbf{t} ,\qquad \mathbf{x} \in \Gamma_i, \: t>0, \label{eq:slip_BCs}
\end{equation}
where $\beta_i$ is the slip length on $\mathbf{x} \in \Gamma_i$ for $i = 0, 1, \ldots N-1$ measured per unit length, 
$\mathbf{t}$ is any unit tangent to the surface such that $\mathbf{t} \cdot \mathbf{n} = 0$, and the superscript $\mathrm{T}$ denotes the transpose \citep{Lauga2007Microfluidics}.   
In the case that $\beta_i = 0$ for some $i$, then the standard no--slip and no--flux boundary conditions for the flow are enforced on $\Gamma_i$. 
We specify initial conditions through
\begin{equation}
\mathbf{v}(\mathbf{x},0) = \mathbf{v}_0(\mathbf{x}), \qquad p(\mathbf{x},0) = p_0(\mathbf{x}), \qquad \mathbf{x} \in \Omega_f,\label{eq:Navier_Stokes_initial_conditions}
\end{equation}
where $\mathbf{v}_0$ and $p_0$ are known functions. We discuss the choice for these in Section~\ref{sec:numerical_methods}.

We denote the  concentration of the solute within the fluid by $c(\mathbf{x},t)$, measured in particle number per unit volume.  Appealing to the conservation of mass and assuming no fluid--phase reactions occur, the spatio--temporal evolution of the solute concentration is given by
\begin{equation}
\frac{\partial c}{\partial t} + \nabla \cdot \left( \mathbf{v} c \right) =D \nabla \cdot \left( \nabla c \right) , \qquad \mathbf{x} \in \Omega_f, t >0, \label{eq:Convection_Diffusion_Equation}
\end{equation}
where $D\geq 0$  is the  solute diffusion coefficient which we assume to be scalar and constant. We assume a known concentration of the solute at the inlet, and prescribe zero flux of the solute at the outlet  and  on the  external walls as follows:
\begin{subequations}
\label{eq:Reaction_diffusion_advection_boudary_conditions}
\begin{empheq}[right={\qquad t > 0,}]{align}
c &= c_{\text{in}}, \qquad & \mathbf{x} &\in \partial \Omega_{\text{in}}, \\
\nabla c \cdot \mathbf{n} &= 0, \qquad &\mathbf{x} &\in\partial \Omega_{\text{out}}    \cup \partial \Omega_{\text{wall}},
\end{empheq}
\end{subequations}
where $c_{\text{in}} > 0$ is assumed to be constant.   

\subsection{Models for surface reactions}
We are required to specify boundary conditions for $c(\mathbf{x},t)$ on  $\mathbf{x} \in \Gamma_i$, to describe the surface reactions occurring here.  In general, there are two stages of adsorption of a particle from the bulk solution to the solid surface. The first stage involves the diffusion of  particles from the bulk solution to the subsurface and the  second stage then involves the transfer of  particles  from the subsurface to the surface.  After the adsorption of a molecule at the interface, there is a reorientation of the colloid molecules at the surface, which results in a change in the surface tension \citep{birdi2008handbook}.  Assuming that both the rate of diffusion of the particle from the bulk to the subsurface, and the rate of the transfer of the particles from the subsurface to the surface are  important  in determining the overall rate of reaction, we use a mixed kinetic--diffusion adsorption description, given by
\begin{equation}
-D \nabla c \cdot \mathbf{n} = \frac{\partial m}{\partial t} = f_i(c, m), \qquad \mathbf{x} \in \Gamma_i. \: t >0, \label{eq:Robin_boundary_condition}
\end{equation}
Here $m(\mathbf{x},t)$  is the surface concentration of the particle under consideration \citep{birdi2008handbook}, measured in units of number per unit surface area, which contrasts with $c(\mathbf{x},t)$ being  measured in units of number per unit volume.   
The function $f_i(c,m)$ describes the kinetics of the  rate of change of the surface concentration  on the $i^{\text{th}}$ reactive boundary for $i=0, \ldots N-1$  \citep{Danov2002Adsorption}. 
Equation~\eqref{eq:Robin_boundary_condition}  states that the change in the surface concentration is equal to the flux across the surface, where the movement from the bulk to the surface is  termed \emph{adsorption}, and movement from the surface to the bulk is  termed \emph{desorption}.
If $\Gamma_i$ is nonreactive, for some $i$, then $f_i = 0$, so the adsorbed concentration on this boundary type remains constant, and a no--flux boundary condition for the solute concentration  is prescribed. For reactive boundaries, the choice of $f_i$ and its dependence on $c$ and $m$ is highly influential in correctly describing the reaction dynamics at the solid--fluid interface. A number of different isotherms exist for describing these dynamics, dependent on the solute attributes, the order of the reaction, and the interface type. 

The simplest of these is the \emph{Henry} isotherm, which assumes a linear relationship  between the surface pressure and the number of adsorbed particles, and takes the form
\begin{equation}
f_i = \kappa_{a,i} c  - \kappa_{d,i} m, \qquad \mathbf{x} \in \Gamma_i, \: t > 0. \label{eq:Henry_Isotherm}
\end{equation}
Here $\kappa_{a,i} \geq 0$ is the rate of adsorption, measured in unit length per unit time, and $\kappa_{d,i} \geq 0$ is the rate of desorption, measured per unit time, at reactive boundary type $i$ for $i = 0, \ldots N-1$.  Equation~\eqref{eq:Henry_Isotherm} states that the rate of adsorption is proportional to the concentration of particles in solution at the reactive surface. As such, the rate of adsorption predicted does not saturate at higher surface concentrations. However, physically we  expect the rate of adsorption to decrease as the quantity of adsorbed particles increases and the available surface area for adsorption decreases.  Even though the Henry isotherm predicts no  limit to surface concentration and does not model any interaction between the particles,  it has been used in a large number of analytical studies  due to its linearity.

The \emph{Langmuir} adsorption isotherm was the first to be derived mathematically, and is suitable to describe the adsorption of a monolayer of  localized non--ionic non--interacting molecules at a 2D solid interface, and a derivation from statistical physics may be found  in \citep{Baret1969Theoretical}.  It is also  frequently used to describe the  adsorption of molecules at a solid--liquid interface, and is described by
\begin{equation}
f_i = \kappa_{a,i} c\left( 1 - \frac{m}{m_{\infty,i}}\right) - \kappa_{d,i} m, \qquad \mathbf{x} \in \Gamma_i, \: t > 0. \label{eq:Langmuir_Isotherm}
\end{equation}
Here $m_{\infty, i}>0$ is the maximal possible adsorbed surface concentration, measured in number per unit area, at reactive boundary type $i$ for $i = 0, \ldots N-1$.   In comparison to the Henry isotherm, the Langmuir isotherm predicts a decrease in the rate of adsorption as the adsorbed concentration increases  due to the reduction in available adsorption surface.  The Henry isotherm, given in Equation~\eqref{eq:Henry_Isotherm}, is a linearization of Equation~\eqref{eq:Langmuir_Isotherm}, explaining why it produces an accurate representation only at low surface concentrations.

More complex descriptions of the reaction kinetics exist to describe non--localized adsorption and particles which interact.  For example, the \emph{Frumkin} isotherm describes localized adsorption where particle interaction is accounted through an additional  parameter:
\begin{equation}
f_i = \kappa_{a,i} c\left( 1 - \frac{m}{m_{\infty,i}}\right) - \kappa_{d,i} m\exp\left( -\displaystyle \frac{2 \beta m}{kT} \right) , \qquad \mathbf{x} \in \Gamma_i, \: t > 0. \label{eq:Frumkin_Isotherm}
\end{equation} 
where   $k$ is the Boltzmann constant and $T$ is the temperature in Kelvin. The parameter $\beta\geq 0 $ describes how cooperative the reaction is, and is related to the interaction energy between the  particles. In the case that $\beta = 0$, the Langmuir isotherm is recovered.    The Frumkin isotherm is infrequently  used in the mathematical modeling of colloidal transport, which may be due its nonlinearity and the difficulties in determining the interaction energy between the particles.

We make the assumption that the adsorption or desorption of our solute does not alter the position of the reactive boundary, which in the case of small volumes of particles being adsorbed is sufficiently accurate. By the conservation of mass,
such an assumption implies any adsorption or desorption on the surface is represented by a
corresponding increase or decrease in the density of the solid material through time. In some cases, for example when the molecules are big or the number being adsorbed is large, interface evolution needs to be considered and may be done in a similar manner to \citet{Tartakovsky2008Hybrid,Roubinet2013Hybrid} and \citet{Boso2013Homogenizability}. This is particularly important in the application of rock dissolution and precipitation, where large geometrical changes are observed.

To close the system of equations, we impose the initial conditions 
\begin{align}
c(\mathbf{x}, 0) = c_0(\mathbf{x}), \qquad \mathbf{x} \in \Omega_f, \qquad m(\mathbf{x},0) = m_{0,i}(\mathbf{x}) , \quad \mathbf{x} \in \Gamma_i, \label{eq:CDR_initial_conditions}
\end{align}
where $c_0$ and $m_{0,i}$ are known for  $i = 0, 1, \ldots N-1$.

Our problem is, therefore, described by two systems of equations with one--way coupling; the incompressible NS equations, described by \eqref{eq:Navier_Stokes_equations}~--~\eqref{eq:Navier_Stokes_initial_conditions} and the CD equation described by \eqref{eq:Convection_Diffusion_Equation}~--~\eqref{eq:Reaction_diffusion_advection_boudary_conditions},  with reactive boundaries conditions  \eqref{eq:Robin_boundary_condition},  initial conditions \eqref{eq:CDR_initial_conditions}, and a  description of the reaction kinetics, for example Equation~\eqref{eq:Henry_Isotherm}, \eqref{eq:Langmuir_Isotherm} or \eqref{eq:Frumkin_Isotherm}.  We now cast  the equations into dimensionless variables, before detailing the methods used to obtain a numerical solution. 

\subsection{Nondimensionalization}
\label{sec:nondimensionalisation}
 Using a caret notation to distinguish the dimensionless variable from its dimensional equivalent, we let 
\begin{align*}
\mathbf{x} = L \hat{\mathbf{x}}, \quad \mathbf{v} = V_{\text{in}} \hat{\mathbf{v}}, \quad t= \frac{L \hat{t}}{V_{\text{in}}} , \quad p = P_{\text{out}} + \rho V_{\text{in}}^2 \hat{p}, \quad 
c = c_{\text{in}} \hat{c}, \quad m = c_{\text{in}} L \hat{m}, \quad M = c_{\text{in}} L^3 \hat{M},
\end{align*}
where $L>0$ is a typical length scale of the computational domain and $V_{\text{in}} = \| \mathbf{V}_{\text{in}} \|$. 
As  our computational domain consists of voxels, the relationship between each voxel and its material property is conserved upon nondimensionalization, while the length, area and volume of each voxel scales with $L$, $L^2$ and $L^3$ respectively.  Given this, we let  $\hat{\Omega}$, $\hat{\Omega}_s$ and $\hat{\Omega}_f$, with boundaries $\partial \hat{\Omega}_{\text{in}}$,  $\partial \hat{\Omega}_{\text{out}}$,  $\partial \hat{\Omega}_{\text{wall}}$ and $\hat{\Gamma}_i$ for $i = 0, 1, \ldots N-1 $ represent the dimensionless versions of the equivalent dimensional domains and boundaries, where the voxel size is scaled accordingly. 
In dimensionless variables, we, therefore, have
\begin{subequations}
\label{eq:Equations_ND}
\begin{empheq}[right={\qquad \hat{\mathbf{x} }\in \hat{\Omega}_f, \: \hat{t} > 0,}]{align}
\left( \displaystyle\frac{\partial \hat{\mathbf{v}} }{\partial \hat{t}} + \hat{\mathbf{v}} \cdot \hat{ \nabla} \hat{\mathbf{v}} \right) &= - \hat{\nabla} \hat{p} + \frac{1}{\mathrm{Re}} \hat{\nabla}^2 \hat{\mathbf{v}} ,\label{eq:Navier_Stokes_equations_eq1_ND} \\
\hat{\nabla} \cdot \hat{\mathbf{v}} &= 0, \label{eq:Navier_Stokes_equations_eq2_ND}\\
\frac{\partial \hat{c}}{\partial \hat{t}} + \hat{\nabla} \cdot \left( \hat{\mathbf{v}} \hat{c} \right) &= \frac{1}{\mathrm{Pe}} \hat{\nabla} \cdot \left( \hat{\nabla } \hat{c} \right) , \label{eq:Convection_Diffusion_Equation_ND}
\end{empheq}
\end{subequations}
where 
\begin{equation}
\mathrm{Re}= \displaystyle\frac{L \rho V_{\text{in}}}{\mu}, \qquad \mathrm{Pe} =\displaystyle \frac{V_{\text{in}} L}{ D}, \label{eq:Reynolds_Peclet}
\end{equation}
are the global Reynolds and P\'eclet numbers respectively, being the ratio between the inertial and viscous forces and the ratio between advective and diffusive transport rates respectively. Boundary conditions are given by 
\begin{subequations}
\label{eq:boundary_conditions_ND}
\begin{empheq}[right={\qquad \hat{t} > 0.}]{align}
\hat{\mathbf{v}} &= \mathbf{n}, \qquad & \hat{\mathbf{x}} &\in \partial \hat{\Omega}_{\text{in}}, \label{eq:NS_BC_ND_1} \\
\hat{p} &= 0 \text{ and } \hat{\nabla} \hat{\mathbf{v}} \cdot \mathbf{n} = \mathbf{0} \qquad & \hat{\mathbf{x}}&\in \partial \hat{\Omega}_{\text{out}},  \label{eq:NS_BC_ND_2}\\
\hat{\mathbf{v}} &= \mathbf{0}, \qquad&\hat{\mathbf{x}} &\in \partial \hat{\Omega}_{\text{wall}}, \label{eq:NS_BC_ND_3} \\
\hat{\mathbf{v}} \cdot \mathbf{n} &= 0, \text{ and } 
\hat{\mathbf{v} }\cdot \mathbf{t} = \hat{\beta}_i \mathbf{n} \cdot \left(\hat{\nabla} \hat{ \mathbf{v}} + \left( \hat{\nabla} \hat{\mathbf{v}} \right)^{\mathrm{T}}\right) \cdot \mathbf{t} ,\qquad &\hat{\mathbf{x}} &\in \hat{\Gamma}_i, \label{eq:Slip_BC_ND} \\
\frac{1}{\mathrm{Pe}} \hat{\nabla} \hat{c} \cdot \mathbf{n} &= \frac{\partial \hat{m}}{\partial \hat{t}} = \hat{f}_i(\hat{c}, \hat{m}), \qquad &\hat{\mathbf{x}} &\in \hat{\Gamma}_i \label{eq:Reactive_BC_ND},\\
\hat{c} &= 1, \qquad &\hat{\mathbf{x}} &\in \partial \hat{\Omega}_{\text{in}}, \label{eq:C_in_ND} \\
\hat{\nabla } \hat{c} \cdot \mathbf{n} &= 0, \qquad &\hat{\mathbf{x}} &\in \partial \hat{\Omega}_{\text{out}} \cup \partial \hat{\Omega}_{\text{wall}} , \label{eq:C_out_ND}
\end{empheq}
\end{subequations}
with $ \hat{\beta}_i= \displaystyle\frac{\beta_i}{L} $. 
In the case of the Henry isotherm, we have
\begin{equation}
\hat{f}_i = \mathrm{Da}^{\textrm{I}}_{a,i} \hat{c}  - \mathrm{Da}^{\textrm{I}}_{d,i} \hat{m}, \qquad \hat{\mathbf{x}} \in \hat{\Gamma}_i, \: \hat{t} > 0,  \label{eq:Henry_ND}
\end{equation}
whereas, nondimensionalization of the Langmuir and Frumkin  isotherms yields
\begin{equation}
\hat{f}_i = \mathrm{Da}^{\textrm{I}}_{a,i} \hat{c}\left( 1 - \frac{\hat{m}}{\hat{m}_{\infty,i}}\right) - \mathrm{Da}^{\textrm{I}}_{d,i} \hat{m}, \qquad \hat{\mathbf{x}} \in \hat{\Gamma}_i, \: \hat{t} > 0,  \label{eq:Langmuir_ND}
\end{equation}
and 
\begin{equation}
\hat{f}_i = \mathrm{Da}^{\textrm{I}}_{a,i} \hat{c}\left( 1 - \frac{\hat{m}}{\hat{m}_{\infty,i}}\right) - \mathrm{Da}^{\textrm{I}}_{d,i} \hat{m} \exp\left( - \hat{\beta} \hat{m}\right) , \qquad \hat{\mathbf{x}} \in \hat{\Gamma}_i, \: \hat{t} > 0,  \label{eq:Frumkin_ND}
\end{equation}
respectively, 
for $i=0, 1, \ldots N-1$, where $\hat{\beta} = \displaystyle \frac{ 2\beta  c_{\text{in}} L }{kT}  $ and $\hat{m}_{\infty,i} =\displaystyle\frac{ m_{\infty,i}}{c_{\text{in}} L}$.  In \eqref{eq:Henry_ND}~--~\eqref{eq:Frumkin_ND}   the Damk\"ohler numbers are given by 
\begin{equation}
\mathrm{Da}_{a,i} = \displaystyle\frac{\kappa_{a,i}}{V_{\text{in}}}, \qquad \text{and} \qquad \mathrm{Da}_{d,i} = \displaystyle\frac{\kappa_{d,i}L}{V_{\text{in}}}, \label{eq:Damkohler_numbers}
\end{equation}
and describe,  for each $i$,  the ratio of the rate of reaction (either adsorptive or desorptive) to the rate of advective transport.  The initial conditions are given by 
\begin{subequations}
\label{eq:Initial_Conditions_ND}
\begin{empheq}[ ]{align}
\hat{\mathbf{v}}(\hat{\mathbf{x}},0) &= \hat{\mathbf{v}}_0(\hat{\mathbf{x}}), \qquad\hat{p}(\hat{\mathbf{x}},0)=\hat{p}_0(\hat{\mathbf{x}}), \quad \hat{c}(\hat{\mathbf{x}},0) = \hat{c}_0(\hat{\mathbf{x}}), &&\hat{\mathbf{x}} \in \hat{\Omega}_f, \label{eq:Initial_Conditions_ND_1}\\ \hat{m}(\hat{\mathbf{x}},0) &= \hat{m}_{0,i}(\hat{\mathbf{x}}) ,& &\hat{\mathbf{x}} \in \hat{\Gamma}_i, \label{eq:Initial_Conditions_ND_2}
\end{empheq}
\end{subequations}
where $\hat{\mathbf{v}}_0(\hat{\mathbf{x}}) = \displaystyle\frac{\mathbf{v}_0(\mathbf{x})}{V_{\text{in}}}$, $\hat{p}_0 (\hat{\mathbf{x}})= \displaystyle\frac{ p_0(\mathbf{x}) - P_{\text{out}}}{\rho V_{\text{in}}^2 } $, $\hat{c}_0(\hat{\mathbf{x}}) = \displaystyle \frac{c_0(\mathbf{x})}{c_{\text{in}}}$ and $\hat{m}_{0,i}(\hat{\mathbf{x}}) = \displaystyle\frac{m_{0,i}(\mathbf{x})}{c_{\text{in}} L}$ for $i=0, 1, \ldots N-1$.

We now proceed to discuss the numerical methods used to obtain an approximate solution to our dimensionless system of equations given by \eqref{eq:Equations_ND}, with boundary conditions given by \eqref{eq:boundary_conditions_ND}~--~\eqref{eq:Langmuir_ND} and initial conditions specified through \eqref{eq:Initial_Conditions_ND}.

\section{Numerical methods}
\label{sec:numerical_methods}
The full system of equations cannot be solved using analytical techniques, and so numerical methods need to be employed to calculate an approximate solution. 
 We employ FV methods to numerically solve our system of equations, motivated by their  local mass conserving properties. Other methods, for example   LB  or finite 
difference methods may also be used for solving the flow problem.  We note that our  CD  solver is completely compatible with LB methods (the compatibility of our solver with  finite difference methods depends on the grid selection). Although the authors are not aware of a detailed 
comparison of the performance of FV  and LB methods  in  solving the NS equations at  the pore--scale,  some 
incomplete internal studies indicate that LB methods can be advantageous for geometries with a very low porosity and a high tortuosity,  while  FV  methods
are favorable in  other cases.
 
Due to the one--way coupling between our two systems of equation,  the velocity and pressure solutions are  at steady state. For the sake of  generality, we consider the unsteady equations, and begin by solving the system of equations describing the fluid flow, namely \eqref{eq:Navier_Stokes_equations_eq1_ND}, \eqref{eq:Navier_Stokes_equations_eq2_ND} along with \eqref{eq:NS_BC_ND_1}~--~\eqref{eq:Slip_BC_ND},  to obtain a steady state numerical solution where  $\displaystyle \frac{\partial \hat{\mathbf{v}}}{\partial \hat{t}} = \mathbf{0}$ is satisfied. This is achieved using a Chorin fractional timestepping method and we refer the reader to \citet{Ciegis2007On} and \citet{Lakdawala2010phd} for full details and for further references, where the methods used are  described. 

Once the solution of $\hat{\mathbf{v}}$ is obtained, we proceed to solve the system of equations describing the solute transport and reaction, \eqref{eq:Convection_Diffusion_Equation_ND} along with the boundary conditions \eqref{eq:Reactive_BC_ND}~--~\eqref{eq:Langmuir_ND} and initial conditions \eqref{eq:Initial_Conditions_ND}, using a FV method with a cell-centered grid. For the sake of brevity, full details of the numerical method employed are not given, but we refer the reader to, for example \citet{causonintroductory}, for more detailed information on  FV methods.  
Firstly, dimensionless time is uniformly partitioned into $Q$ time points, denoted by $\hat{t}^0, \hat{t}^1,\ldots \hat{t}^{Q-1}$ with $\hat{t}^{k} = k \left(\Delta \hat{t}\right)$, where $\Delta \hat{t}$ is the dimensionless timestep size. 
Then the spatial  domain, $\hat{\Omega}$, is  split into $P$ non--overlapping cubic finite volumes, $\mathcal{B}_l$ for $l = 0, 1, \ldots P_1$, which span the 3D computational domain, such that $\hat{\Omega} = \cup_{l=0}^{P-1} \mathcal{B}_l$. Considering a single representative finite volume, $\mathcal{B}_l$, we denote its six faces by $\mathcal{F}_{l,j}$ with center $\hat{\mathbf{x}}_j$ where the subscript $j=e,w,n,s,t,b$ denotes the east, west, north, south, top and bottom faces respectively. Integration of \eqref{eq:Convection_Diffusion_Equation_ND} over the control volume, $\mathcal{B}_l$, and time interval $[\hat{t}^{k}, \hat{t}^{k+1}]$, upon application of the divergence theorem, yields
\begin{align*}
\int_{\mathcal{B}_l} \hat{c}(\hat{\mathbf{x}}, \hat{t}^{k+1}) \: \mathrm{d} V - \int_{\mathcal{B}_l} \hat{c}(\hat{\mathbf{x}}, \hat{t}^{k}) \: \mathrm{d} V+\int_{\hat{t}^{k}}^{\hat{t}^{k+1}}\int_{\partial \mathcal{B}_l} \hat{\mathbf{v}} \hat{c} \cdot \mathbf{n} \: \mathrm{d} S \: \mathrm{d}\tau = \frac{1}{\mathrm{Pe}} \int_{\hat{t}^{k}}^{\hat{t}^{k+1}}\int_{\partial \mathcal{B}_l} \hat{\nabla} \hat{c} \cdot \mathbf{n}\: \mathrm{d} V \: \mathrm{d} \tau,
\end{align*}
where $\partial \mathcal{B}_l$ is the boundary of $\mathcal{B}_l$, so that $\partial \mathcal{B}_l = \sum_{j=e,w,n,s,t,b} \mathcal{F}_{l,j}$ . 
Denoting the center of the finite volume by $\mathbf{x}_c$ and using the approximations $\displaystyle\int_{\mathcal{F}_{l,j}} \phi(\hat{\mathbf{x}},\hat{t}) \: \mathrm{d}S =\hat{A}_j \phi(\hat{\mathbf{x}}_j,\hat{t})$ and $\displaystyle\int_{\mathcal{B}_l} \phi(\hat{\mathbf{x}},\hat{t}) \: \mathrm{d}S =|\mathcal{B}_l| \phi(\hat{\mathbf{x}}_c,\hat{t})$, for some scalar function $\phi$ for $j=n,s,e,w,t,b$, where $\hat{A}_j$ is the area of the face $\mathcal{F}_{l,j}$, we have
\begin{multline*}
|\mathcal{B}_l| \left(\hat{c}(\hat{\mathbf{x}}_c, \hat{t}^{k+1}) - \hat{c}(\hat{\mathbf{x}}_c, \hat{t}^{k})\right)\\+
\int_{\hat{t}^{k}}^{\hat{t}^{k+1}} \left( \hat{A}_e \left[\hat{\mathbf{v}} \hat{c}\right]_{\hat{\mathbf{x}}_e } -\hat{A}_w \left[\hat{\mathbf{v}} \hat{c}\right]_{\hat{\mathbf{x}}_x } +\hat{A}_n \left[\hat{\mathbf{v}} \hat{c}\right]_{\hat{\mathbf{x}}_n } - \hat{A}_s\left[\hat{\mathbf{v}} \hat{c}\right]_{\hat{\mathbf{x}}_s } + \hat{A}_t\left[\hat{\mathbf{v}} \hat{c}\right]_{\hat{\mathbf{x}}_t} - \hat{A}_b \left[\hat{\mathbf{v}} \hat{c}\right]_{\hat{\mathbf{x}}_b } \right) \: \mathrm{d}\tau \\ = \frac{1}{\mathrm{Pe}} \int_{\hat{t}^{k}}^{\hat{t}^{k+1}} \left(\hat{A}_e\left[\frac{\partial \hat{c}}{\partial \hat{x}}\right]_{\hat{\mathbf{x}}_e } - \hat{A}_w \left[ \frac{\partial \hat{c}}{\partial \hat{x}}\right]_{\hat{\mathbf{x}}_w } +\hat{A}_n\left[ \frac{\partial \hat{c}}{\partial \hat{y}}\right]_{\hat{\mathbf{x}}_n } -\hat{A}_s \left[ \frac{\partial \hat{c}}{\partial \hat{y}}\right]_{\hat{\mathbf{x}}_s } +\hat{A}_t \left[ \frac{\partial \hat{c}}{\partial \hat{z}}\right]_{\hat{\mathbf{x}}_t } - \hat{A}_b\left[ \frac{\partial \hat{c}}{\partial \hat{z}}\right]_{\hat{\mathbf{x}}_b }\right) \: \mathrm{d}\tau.
\end{multline*}
Denoting $\hat{c}_j(\tau) = \hat{c}(\hat{\mathbf{x}}_j, \tau)$ and $\hat{\mathbf{v}}_j(\tau) = \hat{\mathbf{v}}(\hat{\mathbf{x}}_j, \tau)$ for 
$j=n,s,e,w,t,b,c$, by first order finite difference methods we have
\begin{multline}
|\mathcal{B}_l| \left( \hat{c}_c( \hat{t}^{k+1}) -\hat{c}_c( \hat{t}^{k})\right)+ 
\int_{\hat{t}^{k}}^{\hat{t}^{k+1}} \hat{A}_e\hat{\mathbf{v}}_e \hat{c}_e- \hat{A}_w \hat{\mathbf{v}}_w \hat{c}_w + \hat{A}_n \hat{\mathbf{v}}_n \hat{c}_n- \hat{A}_s \hat{\mathbf{v}}_s \hat{c}_s+ \hat{A}_t \hat{\mathbf{v}}_t \hat{c}_t- \hat{A}_b \hat{\mathbf{v}}_b \hat{c}_b \: \mathrm{d} \tau \\ = \int_{\hat{t}^{k}}^{\hat{t}^{k+1}} \frac{2}{\mathrm{Pe}} \left( \hat{A}_e \frac{\hat{c}_e -\hat{c}_c}{\delta \hat{x}} - \hat{A}_w\frac{\hat{c}_c-\hat{c}_w }{\delta \hat{x}} +\hat{A}_n \frac{\hat{c}_n - \hat{c}_c}{\delta \hat{y}} - \hat{A}_s \frac{\hat{c}_c-\hat{c}_s }{\delta \hat{y}} + \hat{A}_t \frac{\hat{c}_t - \hat{c}_c}{\delta \hat{z}} - \hat{A}_b\frac{\hat{c}_c-\hat{c}_b }{\delta \hat{z}}\right) \: \mathrm{d}\tau, \label{eq:Finite_Volume_Dis_1}
\end{multline}
where $\delta \hat{x}$, $ \delta \hat{y}$  and $ \delta \hat{z}$ are the width, length and height of the control volume $\mathcal{B}_l$. 
By virtue of using voxelized geometry, we know that $\delta \hat{x} = \delta \hat{y} = \delta \hat{z}$, $\hat{A}_j = \left( \delta \hat{x}\right)^2$ for all $j=n,s,e,w,t,b$, and $|\mathcal{B}_l| = \left( \delta \hat{x}\right)^3$. By the implicit Euler method, we, therefore, have
\begin{multline}
\frac{\delta \hat{x} \left(\hat{c}_c^{k+1} - \hat{c}_c^{k}\right)}{ \Delta \hat{t}}
+ \hat{\mathbf{v}}_e^{k+1} \hat{c}_e^{k+1}- \hat{\mathbf{v}}_w \hat{c}_w^{k+1} + \hat{\mathbf{v}}_n^{k+1} \hat{c}_n^{k+1}- \hat{\mathbf{v}}_s^{k+1} \hat{c}_s^{k+1}+ \hat{\mathbf{v}}_t^{k+1} \hat{c}_t^{k+1}- \hat{\mathbf{v}}_b^{k+1} \hat{c}_b^{k+1} \\ = \frac{2 }{\mathrm{Pe} \: \delta \hat{x}} \left( \hat{c}_e^{k+1} + \hat{c}_w^{k+1} +\hat{c}_n^{k+1} + \hat{c}_s^{k+1} + \hat{c}_t^{k+1}+ \hat{c}_b^{k+1} -6 \hat{c}_c^{k+1}\right), \label{eq:Finite_Volume_Dis_2}
\end{multline}
where  $\hat{c}_j^{k}= \hat{c}(\hat{\mathbf{x}}_j, \hat{t}^{k})$ for $j=n,s,e,w,t,b,c$.   In the case that one of the faces of the control volume lies on a boundary, the appropriate boundary conditions must be  discretized; for the inlet, outlet and solid (or symmetry) boundaries this is straightforward due to the Dirichlet and zero Neumann boundary condition imposed there via \eqref{eq:C_in_ND} and \eqref{eq:C_out_ND}. Therefore, we omit the details  for the discretization of the boundary conditions on the external boundary $\partial \hat{\Omega}$.  The appropriate discretization of the reactive boundary conditions, prescribed on the fluid--solid interface through \eqref{eq:Reactive_BC_ND}   and the corresponding description of the  reaction kinetics, here either \eqref{eq:Henry_ND},   \eqref{eq:Langmuir_ND} or \eqref{eq:Frumkin_ND},  is slightly more involved and deserves a more detailed discussion.

In a fully implicit and coupled discretization the resulting discrete equations are  nonlinear  and the Newton method needs to be used  \citep{Kelley1995Iterative}.  In a broad class of practically interesting problems we have considered to date, we have not faced very strong coupling between the dissolved and adsorbed concentrations. Therefore, a fully implicit and coupled discretization was not required and we have found that an operator splitting approach, or just a Picard linearization, has worked well. In this approach, the dissolved particle concentration is computed at  $\hat{t}=\hat{t}^{k+1/2}$, and then the value is  used  to compute the  deposited mass at the time $\hat{t}=\hat{t}^{k+1}$. 
Runge--Kutta methods, or other methods for numerically solving stiff ODEs, are also straightforward to 
implement, and may be the 
subject of future studies if required \citep{Kelley1995Iterative}.  
In order to illustrate the method used in Pore--Chem, we describe the  discretization for the Langmuir isotherm, which is achieved as follows. Firstly \eqref{eq:Langmuir_ND} is substituted into \eqref{eq:Reactive_BC_ND}, which is then split into a Robin boundary condition and an ordinary differential equation:
\begin{subequations}
\label{eq:boundary_conditions_reactive_split}
\begin{empheq}[right={\qquad \hat{\mathbf{x}}\in \hat{\Gamma}_i, \: \hat{t}>0.}]{align}
-\frac{1}{\mathrm{Pe}} \hat{\nabla} \hat{c}\cdot \mathbf{n}  &= \mathrm{Da}^{\textrm{I}}_{a,i} \hat{c}\left( 1 - \frac{\hat{m}}{\hat{m}_{\infty,i}}\right) -\mathrm{Da}^{\textrm{I}}_{d,i} \hat{m}, \label{eq:Robin_part} \\
\frac{\partial \hat{m}}{\partial \hat{t}} &=\mathrm{Da}^{\textrm{I}}_{a,i} \hat{c}\left( 1 - \frac{\hat{m}}{\hat{m}_{\infty,i}}\right) - \mathrm{Da}^{\textrm{I}}_{d,i} \hat{m} . \label{eq:ODE}
\end{empheq}
\end{subequations}
If $\mathrm{Da}^{\textrm{I}}_{a,i}\displaystyle\frac{ \hat{c} }{\hat{m}_{\infty,i}}  +\mathrm{Da}^{\textrm{I}}_{d,i}=0$, then either $\hat{c} =\mathrm{Da}^{\textrm{I}}_{d,i}= 0$ or $\mathrm{Da}^{\textrm{I}}_{a,i} = \mathrm{Da}^{\textrm{I}}_{d,i}=0$. In both of these cases, by \eqref{eq:Langmuir_ND},  $\hat{f}_i =0$  and so no reactions occur at the spatiotemporal point under consideration, in which case, by \eqref{eq:Robin_part}, we have $\hat{\nabla} \hat{c} \cdot \mathbf{n} = 0$ and a zero Neumann boundary condition,  which is straightforward to implement. 
Otherwise, if $\mathrm{Da}^{\textrm{I}}_{a,i} \displaystyle\frac{ \hat{c} }{\hat{m}_{\infty,i}} + \mathrm{Da}^{\textrm{I}}_{d,i}>0$,  assuming that $\hat{c}(\hat{\mathbf{x}},\hat{t})$ is constant over the time period in question, namely $\hat{t} \in [\hat{t}^{k}, \hat{t}^{k+1}]$,  and equal to $\hat{c}(\hat{\mathbf{x}})$ at each spatial point, \eqref{eq:ODE} may be   integrated to give 
\begin{align}
\hat{m}(\hat{\mathbf{x}}, \hat{t}^{k+1}) =\frac{ \mathrm{Da}^{\textrm{I}}_{a,i} \hat{c}(\hat{\mathbf{x}}) - B\exp\left(-\left(\mathrm{Da}^{\textrm{I}}_{a,i} \hat{c}(\hat{\mathbf{x}}) \hat{m}_{\infty,i}^{-1} +\mathrm{Da}^{\textrm{I}}_{d,i} \right) \hat{t}^{k+1} \right) }{ \mathrm{Da}^{\textrm{I}}_{a,i} \hat{c}(\hat{\mathbf{x}}) \hat{m}_{\infty,i}^{-1} + \mathrm{Da}^{\textrm{I}}_{d,i}}, \qquad \hat{\mathbf{x}} \in \hat{\Gamma}_i, \: \hat{t}>0. \label{eq:ODE_exact_Langmuir}
\end{align}
Here  $B$ is a constant of integration which may be evaluated at $\hat{t}=\hat{t}^{k}$ to give
\begin{align}
B = \left( \mathrm{Da}^{\textrm{I}}_{a,i} \hat{c}(\hat{\mathbf{x}}) \left( 1 - \hat{m}(\hat{\mathbf{x}}, \hat{t}^{k}) \hat{m}_{\infty,i}^{-1}\right) -\mathrm{Da}^{\textrm{I}}_{d,i} \hat{m}(\hat{\mathbf{x}}, \hat{t}^{k}) \right)\exp\left(\left(\mathrm{Da}^{\textrm{I}}_{a,i} \hat{c}(\hat{\mathbf{x}}) \hat{m}_{\infty,i}^{-1} +\mathrm{Da}^{\textrm{I}}_{d,i}\right) \hat{t}^{k} \right) . \label{eq:M_exact}
\end{align}
Upon substitution into  \eqref{eq:ODE_exact_Langmuir}, we have
{\small
\begin{multline}
\hat{m}(\hat{\mathbf{x}},\hat{t}^{k+1}) \\= \frac{ \mathrm{Da}^{\textrm{I}}_{a,i} \hat{c}(\hat{\mathbf{x}},\hat{t}^{k}) - \left( \mathrm{Da}^{\textrm{I}}_{a,i}\hat{c}(\hat{\mathbf{x}}, \hat{t}^{k}) \left( 1 - \hat{m}(\hat{\mathbf{x}}, \hat{t}^{k})\hat{m}_{\infty,i}^{-1} \right) - \mathrm{Da}^{\textrm{I}}_{d,i} \hat{m}(\hat{\mathbf{x}}, \hat{t}^{k}) \right)\exp\left( -\left( \mathrm{Da}^{\textrm{I}}_{a,i} c(\hat{\mathbf{x}}, \hat{t}^{k}) \hat{m}_{\infty,i}^{-1} +\mathrm{Da}^{\textrm{I}}_{d,i}\right) (\Delta \hat{t}) \right)}{\mathrm{Da}^{\textrm{I}}_{a,i} \hat{c}(\hat{\mathbf{x}}, \hat{t}^{k}) \hat{m}_{\infty,i}^{-1} + \mathrm{Da}^{\textrm{I}}_{d,i}}, \label{eq:Solution_to_M}
\end{multline}}
for $\hat{\mathbf{x}} \in \hat{\Gamma}_i$, $\hat{t}>0$, 
where we have approximated $\hat{c}(\hat{\mathbf{x}})$ by $\hat{c}(\hat{\mathbf{x}}, \hat{t}^{k})$. This is done to prevent nonlinear terms in unknown variables from appearing in the discretized version of \eqref{eq:Reactive_BC_ND}.  Discretization of the other two isotherms is implemented in a similar manner. 

Consequently, we may approximate the Robin boundary condition, \eqref{eq:Reactive_BC_ND}, on the reactive face \mbox{$\mathcal{F}_{l,j} \in \hat{\Gamma}_i$} for $j=e,w,n,s,t,b$ and $i = 0, 1, \ldots N-1$ using finite difference methods fully implicitly as follows:
{\small
\begin{multline}
-\left( \frac{2n }{\mathrm{Pe} \: \delta \hat{x}}+\mathrm{Da}^{\textrm{I}}_{a,i} \right) c_j^{k+1} +2 n\frac{ c_c^{k+1} }{ \mathrm{Pe} \: \delta \hat{x}} \\= - \left( \frac{\mathrm{Da}^{\textrm{I}}_{a,i} }{\hat{m}_{\infty,i}}+ \mathrm{Da}^{\textrm{I}}_{d,i} \right) \left( \frac{ \mathrm{Da}^{\textrm{I}}_{a,i} \hat{m}_j^{k}- \left(\mathrm{Da}^{\textrm{I}}_{a,i} c_j^{k}\left( 1 - \hat{m}_j^{k}\hat{m}_{\infty,i}^{-1} \right) - \mathrm{Da}^{\textrm{I}}_{d,i} \hat{m}_j^{k}\right)\exp\left( -\left( \mathrm{Da}^{\textrm{I}}_{a,i} c_m^{p }\hat{m}_{\infty,i}^{-1} +\mathrm{Da}^{\textrm{I}}_{d,i}\right) (\Delta t) \right)}{ \mathrm{Da}^{\textrm{I}}_{a,i} c_j^{k} \hat{m}_{\infty,i}^{-1} + \mathrm{Da}^{\textrm{I}}_{d,i}}\right), \label{eq:Discretised_version_Robin_BC}
\end{multline}}
where $n= \pm 1$ is the direction of the outward pointing normal.  
 Using \eqref{eq:Discretised_version_Robin_BC}, the appropriate terms are assembled, along with \eqref{eq:Finite_Volume_Dis_2} minus the relevant diffusive flux term, for the finite volume on which the reactive surface lies into a matrix $\mathcal{A}^{k+1}$ and vector $\mathbf{g}^{k+1}$, where \mbox{$\mathcal{A}^{k+1}\hat{ \mathbf{c}}^{k+1} = \mathbf{g}^{k+1}$} and $\hat{\mathbf{c}}^{k+1}$ is a vector of the dimensionless solutions $\hat{c}(\hat{\mathbf{x}},\hat{t}^{k+1})$ at the discretized points of the computational domain.  Once all the terms for all the finite volumes within the domain have been assembled into $\mathcal{A}^{k+1}$ and $\mathbf{g}^{k+1}$, the equation $\mathcal{A}^{k+1} \hat{\mathbf{c}}^{k+1}= \mathbf{g}^{k+1}$ is solved using a biconjugate gradient stabilized method to give the updated fluid concentration, $\hat{c}^{k+1}$, at each discrete spatial point in $\hat{\Omega}_f$, while the updated adsorbed concentration, $\hat{m}^{k+1}$, is given by \eqref{eq:M_exact}. Time is then updated, the next timestep considered, and we proceed in the usual manner until the final time point is reached.

\begin{figure}
\footnotesize
\centering
\tikzstyle{decision} = [diamond, draw, fill=white!20, 
    text width=7em, text badly centered, node distance=3cm, inner sep=0pt]
\tikzstyle{startstop} = [diamond, draw, fill=purple!60, 
    text width=7em, text badly centered, node distance=3cm, inner sep=0pt]
\tikzstyle{flowblock} = [rectangle, draw, fill=green!20, 
    text width=9em, text centered, rounded corners, minimum height=4em]
\tikzstyle{generalblock} = [rectangle, draw, fill=purple!60, 
    text width=12em, text centered, rounded corners, minimum height=4em]
\tikzstyle{efficiencyblock} = [rectangle, draw, fill=blue!20, 
    text width=12em, text centered, rounded corners, minimum height=4em]
\tikzstyle{line} = [draw, -latex']
\tikzstyle{inputfiles} = [draw, ellipse,fill=red!20, text width = 7em, text centered, node distance=3cm,
    minimum height=2em]
\tikzstyle{outputfile} = [draw, ellipse,fill=yellow!20, text width =7em, text centered, node distance=3cm,
    minimum height=2em]

\begin{tikzpicture}[node distance =2cm, auto]
   \node [startstop] (Mainsim) {Start simulation};
   \node [inputfiles, left of =Mainsim, node distance = 5cm] (Inputfile){Input file  (description of problem, boundaries, parameters)};
   \node [inputfiles, below left of =Inputfile, node distance =3cm] (Geometryfile){Geometry File};
   \node [generalblock, below right of =Inputfile, node distance =4cm] (Imagedata){Process geometry};
 \node [decision, below of=Imagedata] (DimorNondim) {Solving dimensional problem?};
   \node [generalblock, left of =DimorNondim, node distance =5cm, text width = 12em] (Nondiensionalisation){Cast problem (geometry \& parameters) into dimensionless equivalents};
   \node [generalblock,  below  of = Nondiensionalisation, node distance = 3.5cm] (BoundaryAttributes) {Set attributes for boundaries};
   \node [flowblock,   right of = BoundaryAttributes, node distance = 6cm] (Flowsolve) {Solve fluid flow, $\hat{\mathbf{v}}, \hat{p} \in \hat{\Omega}_f$  (Chorin method with fractional timestepping until $  \displaystyle\frac{\hat{\partial} \hat{\mathbf{v}}}{\hat{\partial} \hat{t}}< \text{tol}$ is achieved};
   \node [outputfile, above right of = Flowsolve, node distance = 5cm ] (OutputFlow) {Export flow solution visualization file};
    \node [efficiencyblock,  right of = Flowsolve, node distance = 5cm] (Tloop) {Start time loop ($k=0, \hat{t}^k=0$)}; 
    \node [decision, below  of=DimorNondim, node distance = 7cm] (endtime) {Has end time been reached? (Is $\hat{t}^k \geq \hat{T}_{\text{end}}$?)};
       \node [efficiencyblock, left of = endtime, node distance = 5cm] (FluidConcAss) {Assemble matrix $\mathcal{A}^{k+1}$ and right--hand--side vector $\mathbf{g}^{k+1}$ for the fluid concentration, including Robin Boundary conditions on reactive boundaries};
    \node [efficiencyblock, below  of = FluidConcAss, node distance = 3cm] (SolveFluid) {Solve $\mathcal{A}^{k+1} \mathbf{c}^{k+1} = \mathbf{g}^{k+1}$ using an iterative linear solver to give $\hat{\mathbf{c}}^{k+1}$, the vector of solutions for the fluid concentration , $\hat{c} \in \hat{\Omega}_f$};
    \node [efficiencyblock, below of = SolveFluid, node distance = 3cm] (UpdateRetained) {Solve the reactive ODE to obtain the retained concentration, $\hat{m}$ for $\hat{\mathbf{x}} \in \hat{\Gamma}_i $ };
    \node [decision,  right of=UpdateRetained, node distance = 5cm] (savesolution) {Is current time $=$ SaveStep time?};
    \node [outputfile, right of =savesolution, node distance = 6cm, text width = 6em] (output_vtk) {Export efficiency solution visualization file   };
    \node [efficiencyblock, right of = SolveFluid, node distance = 5cm](updatetime){Update time \\($k = k+1$, $\hat{t}^{k} = \hat{t}^{k-1} + \Delta t$)}; 
 \node [startstop,  right of = updatetime, node distance = 5cm, text width = 4em ] (endprocess) {Stop}; 
   \path[line](Inputfile)--(Imagedata);
    \path [line] (Flowsolve) -- (Tloop);
     \path [line] (FluidConcAss) -- (SolveFluid);
     \path [line] (SolveFluid) -- (UpdateRetained);
    \path [line] (UpdateRetained) -- (savesolution);
    \path [line, dashed] (endtime) -- node {yes} (endprocess);
    \path [line, dashed] (endtime) -- node {no}(FluidConcAss);
    \path [line] (Tloop) -- (endtime);
\path [line] (Flowsolve) -- (OutputFlow);
    \path [line] (updatetime) -- (endtime);
    \path [line] (UpdateRetained) -- (updatetime);
   \path [line, dashed] (savesolution) -- node {yes} (output_vtk);
  \path [line, dashed] (savesolution) -- node {no} (updatetime);
 \path [line] (output_vtk) -- (updatetime);
  \path [line] (BoundaryAttributes) -- (FluidConcAss);
  \path [line] (Inputfile) -- (Geometryfile);
  \path [line] (Mainsim) -- (Inputfile); 
  \path [line] (Geometryfile) -- (Imagedata);
\path [line] (Imagedata) -- (DimorNondim);
 \path [line, dashed] (DimorNondim) -- node {yes} (Nondiensionalisation);
 \path [line, dashed] (DimorNondim) -- node {no} (BoundaryAttributes);
\path [line] (Nondiensionalisation) -- (BoundaryAttributes);
\path [line] (BoundaryAttributes) -- (Flowsolve);

\end{tikzpicture}
\caption{\footnotesize Flow chart to illustrate how numerical computation of the transport with surface reactions is implemented within Pore--Chem, with main features indicated only. The light red ovals indicate input files, while the yellow ovals indicate output files. The green and  the blue rectangular boxes indicate steps involved for  the flow  and efficiency solvers respectively, while the dark red boxes indicate steps involved in both.  The white diamonds indicate decision making steps. In the case that the dimensional system of equations is being solved,  the appropriate dimensionless quantities of interest are replaced by the dimensional equivalent.   }
\label{fig:Flowchart}
\end{figure}
\normalsize

Numerical simulation of the equations is performed in Pore--Chem  and a schematic, outlining the numerical algorithm used,  is given in  Figure~\ref{fig:Flowchart}. In the following  section, the dimensionless equations are solved, but we present results in dimensional quantities.

\section{Results}
\label{sec:Results}

\begin{figure}
\centering
\begin{subfigure}[c]{0.8\textwidth}
\includegraphics[width =  \textwidth]{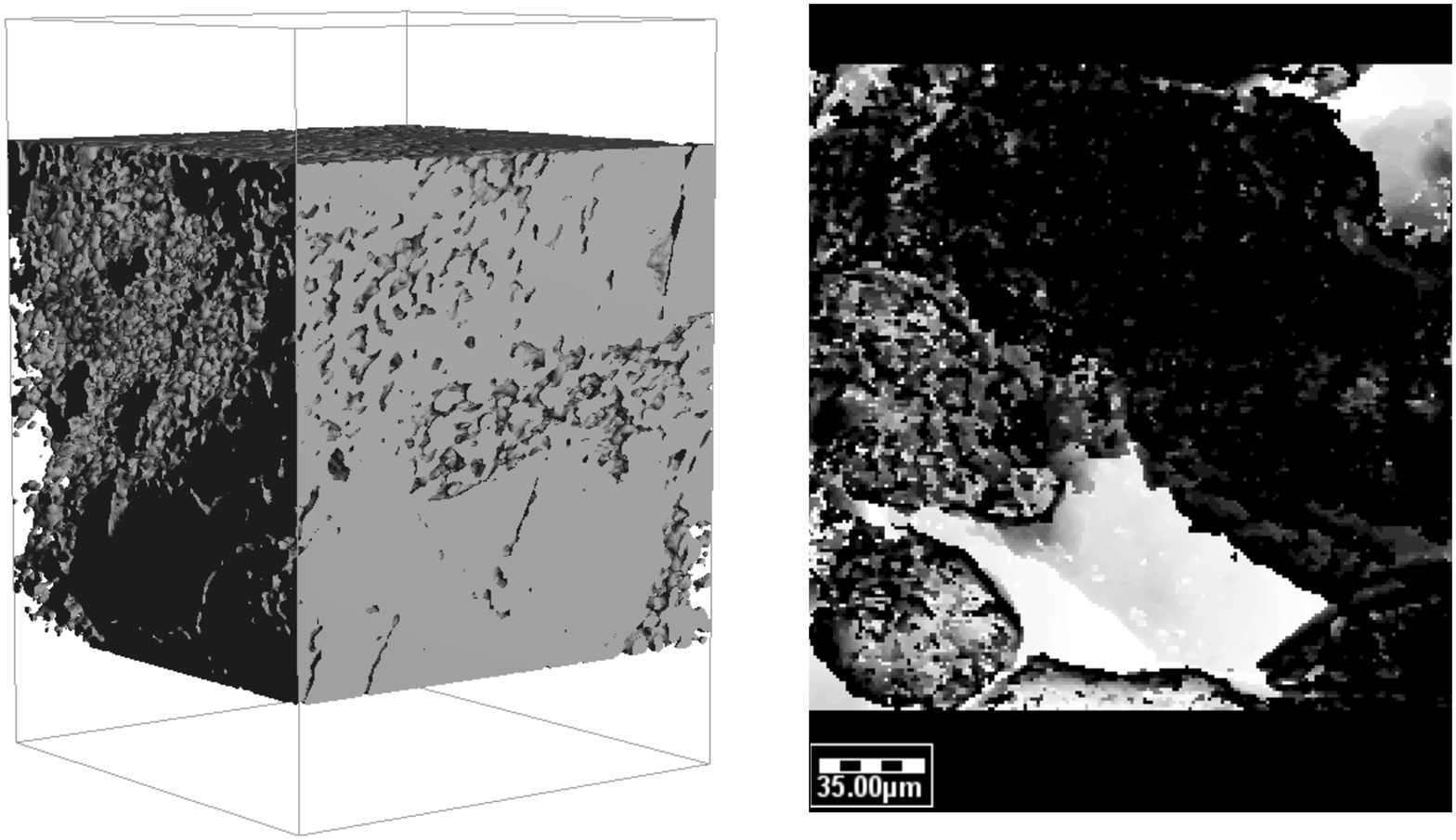} 
\caption{Rock geometry}
\label{fig:Rock_geometry}
\end{subfigure}
\\
\begin{subfigure}[c]{0.8\textwidth}
\includegraphics[width =\textwidth]{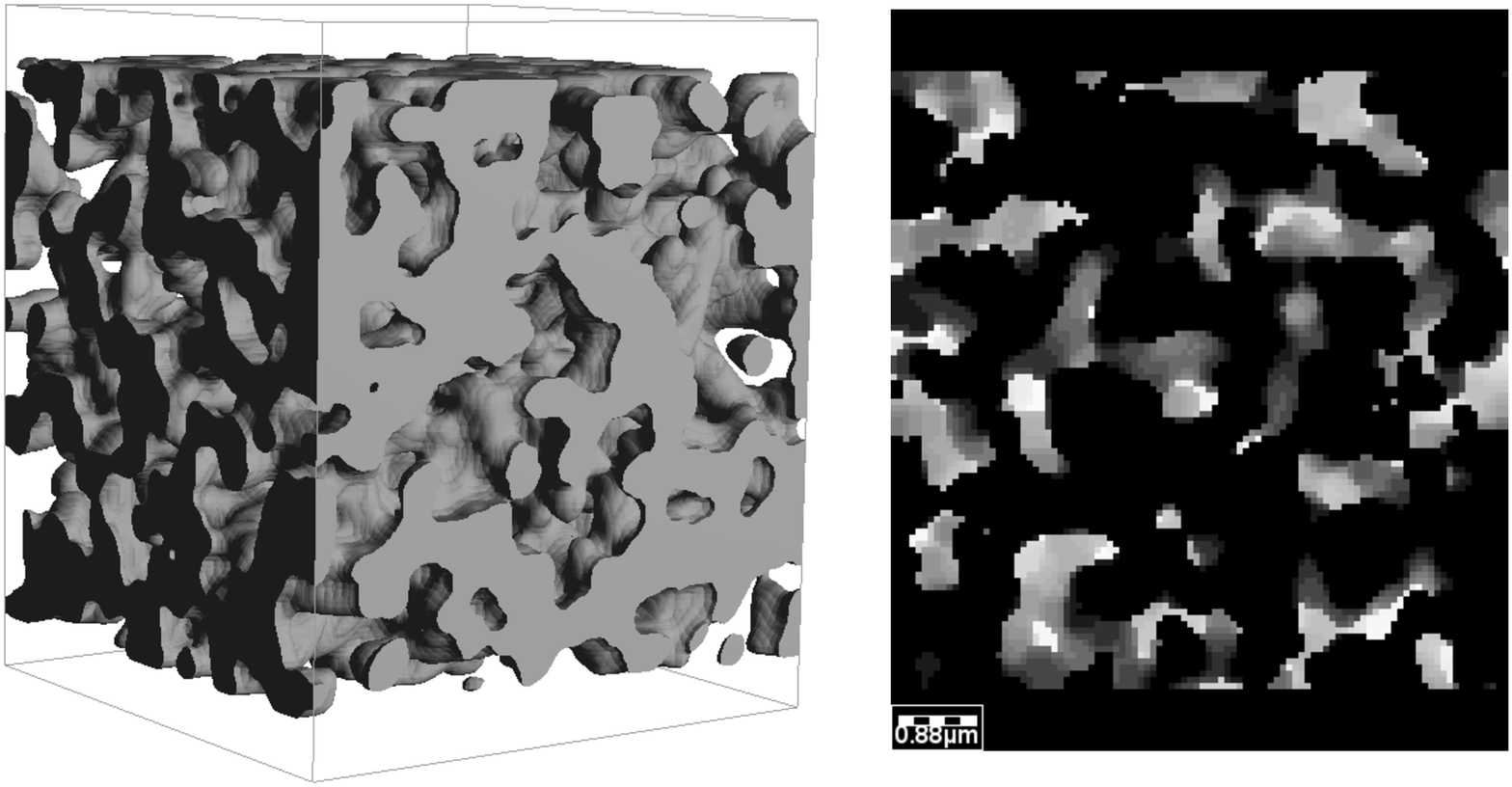}
\caption{Membrane geometry}
\label{fig:Membrane_geometry}
\end{subfigure}
\caption{The two computational domains under consideration, plotted in 3D on the left--hand side of the figure, and in 2D through a representative cross section on the right--hand side of the figure.  Figure~\subref{fig:Rock_geometry} shows the  palatine sandstone geometry obtained through $\mu$--CT,  with a voxel size of $1.4 \times 10^{-6}$~$\mathrm{m}$, as used in \citet{Becker2011Combining}. Figure~\subref{fig:Membrane_geometry} shows a virtually generated geometry which aims to reproduce the morphology of a commercially available microfiltration membrane, with a voxel length of $7.03\times 10^{-8}$~$\mathrm{m}$ (3 s.f.)  This geometry was created using the software GeoDict \citep{Geodict}, and a comparison to experimentally evaluated quantities is made in \citet{Nicolo2014Virtual}.  }  
\label{fig:Computational_geometries}
\end{figure}

We present illustrative results, using the numerical method outlined in Section~\ref{sec:numerical_methods}, on  two separate computational domains, a real geometry and a virtually generated geometry, for two applications where surface reactions are important. The first set of  simulations are performed on a portion of  palatine sandstone, obtained using micro computerized tomography ($\mu$--CT) by Frieder Enzmann at the Johannes Gutenberg University of Mainz  \citep{Becker2011Combining}. Surface reactions are highly important in many 
fields of the Earth sciences, including calcite growth, oxidation--reduction reactions, formation of 
biofilms, to name only a few \citep{Steefel2005Reactive}.  The second set of results is performed  on a
computational domain virtually constructed to be representative of a commercially available microfiltration functionalized membrane. The use of functionalized 
membranes is a promising method for removing contaminants from water, and involves treating the pore walls of the membrane so that they adsorb  certain microorganisms  
or drugs \citep{Ulbricht2006Advanced}.  Such membranes have pore sizes on the sub--micron scale and    the resolution provided by $\mu$--CT imaging techniques is not high enough to give representative images, motivating the use of a virtually generated  geometry. The two computational domains under consideration are shown in Figure~\ref{fig:Computational_geometries}.

\begin{figure}
\centering
\includegraphics[width = 0.4\textwidth]{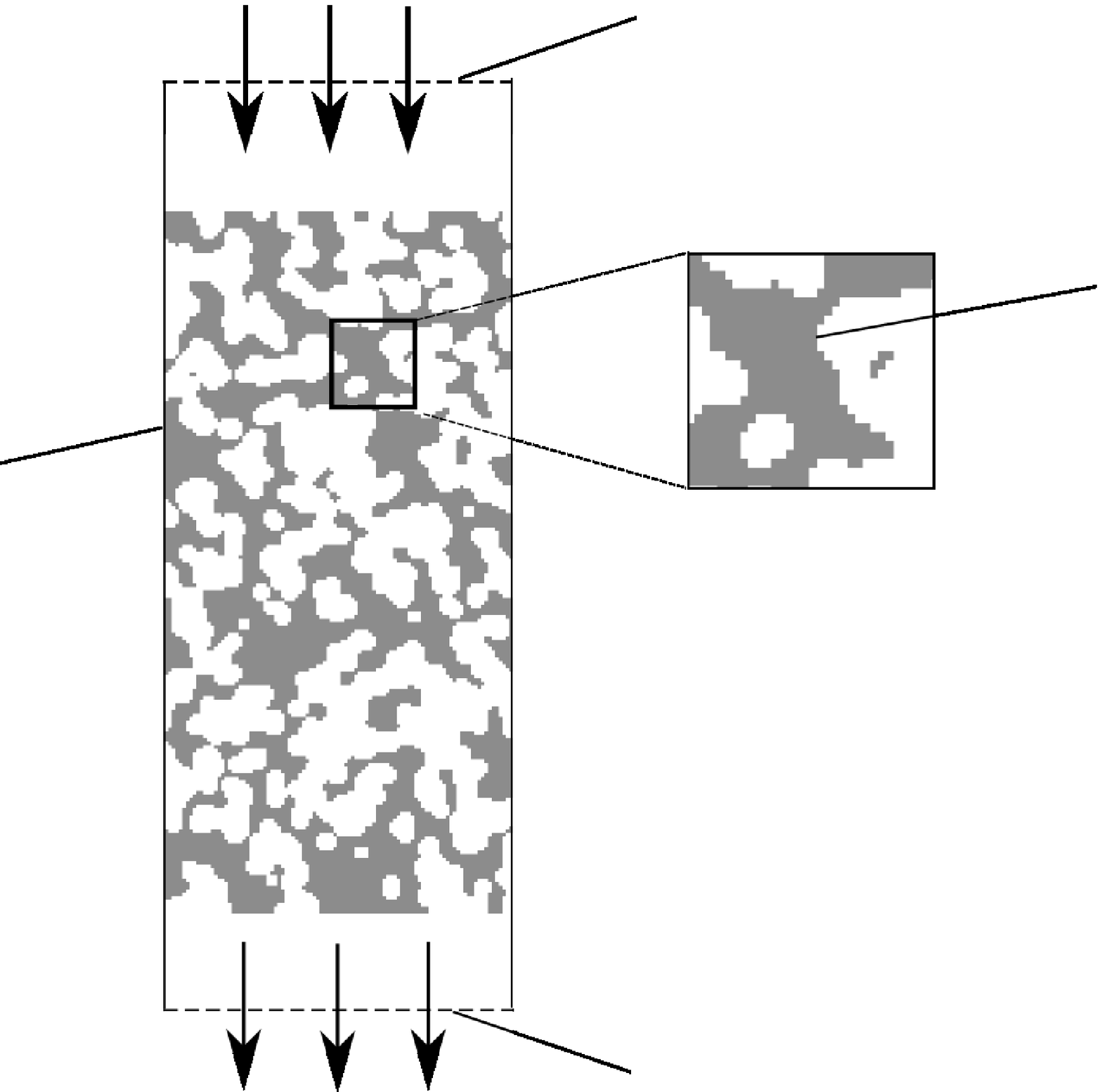}
\put(-70, -2){\small $\partial \Omega_{\text{out}}$}
\put(-70, 175){\small $\partial \Omega_{\text{in}}$}
\put(3, 130){\small $\Gamma$}
\put(-213, 100){\small $\partial \Omega_{\text{wall}}$}
\caption{Schematic to illustrate the computational domain. The solid microfiltration membrane is show in grey while the water is shown in white. Voxels are added to the top and bottom of the domain, at the inlet and outlet, to enable free--flow to develop. Modified, with permission, from \citet{Nicolo2014Virtual}. }
\label{fig:DeadEndSetup}
\end{figure}

For the numerical simulations presented, a dead--end setup is used, with a schematic illustrating the domain and boundary conditions shown in Figure~\ref{fig:DeadEndSetup}.  Layers of pure water voxels are added at the inlet and at the outlet, as shown in Figure~\ref{fig:Computational_geometries} and illustrated in Figure~\ref{fig:DeadEndSetup},  to allow free flow to develop. This results in a total computational domain size of $200 \times 200 \times 300$ voxels for the sandstone geometry, and $100 \times 100 \times 120$ voxels for the membrane geometry. 
For the results presented one type of reactive solid--fluid interface is considered, so that $N=1$, and consequently we now drop the $i$ subscript notation relating to the type of reactive boundary.

\subsection{Fluid flow}
\begin{figure}
\centering
\begin{subfigure}[c]{0.495\textwidth}
\includegraphics[width = \textwidth]{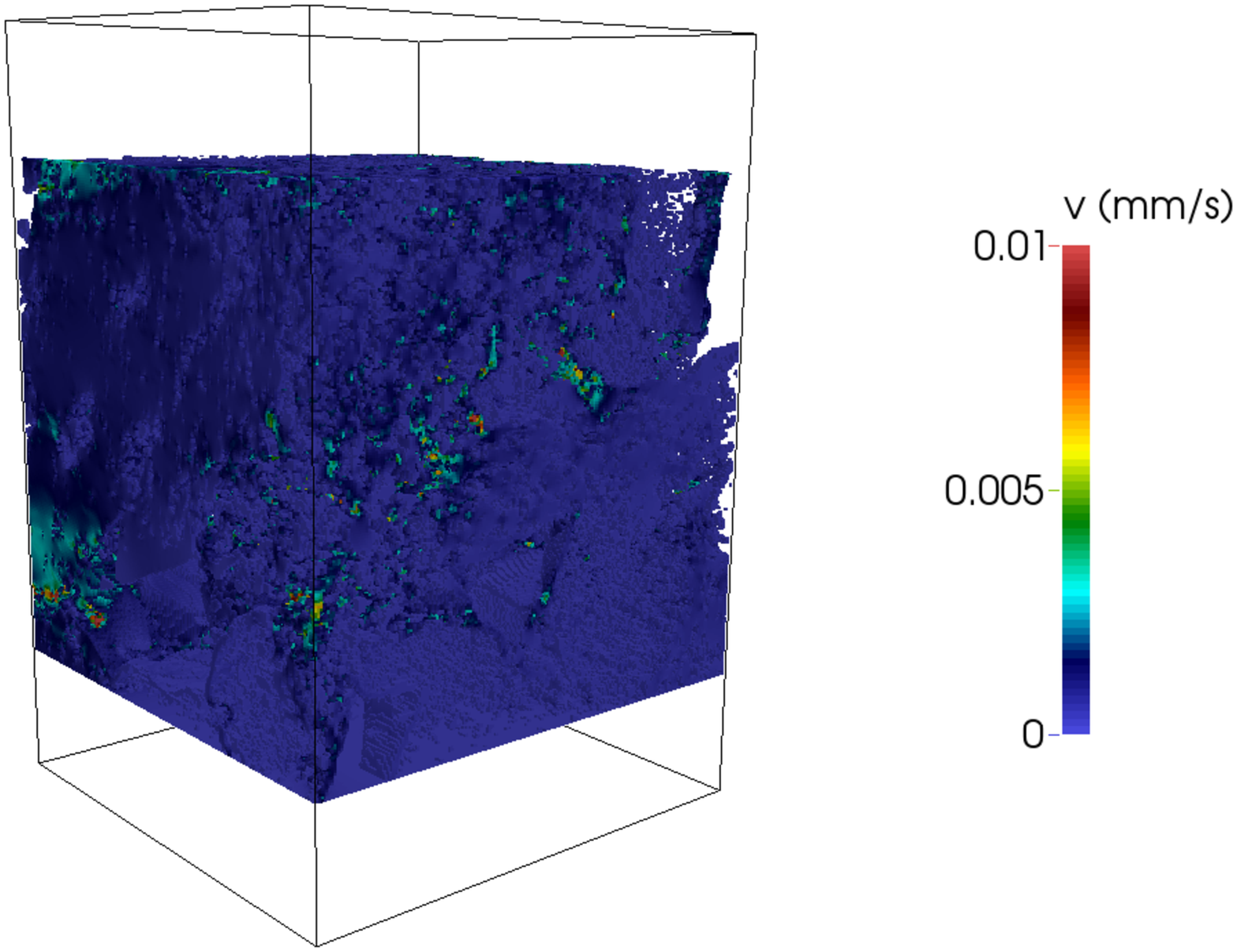}
\caption{Velocity magnitude, rock}
\label{fig:Rock_velocity_rescaled}
\end{subfigure}
\begin{subfigure}[c]{0.495\textwidth}
\includegraphics[width = \textwidth]{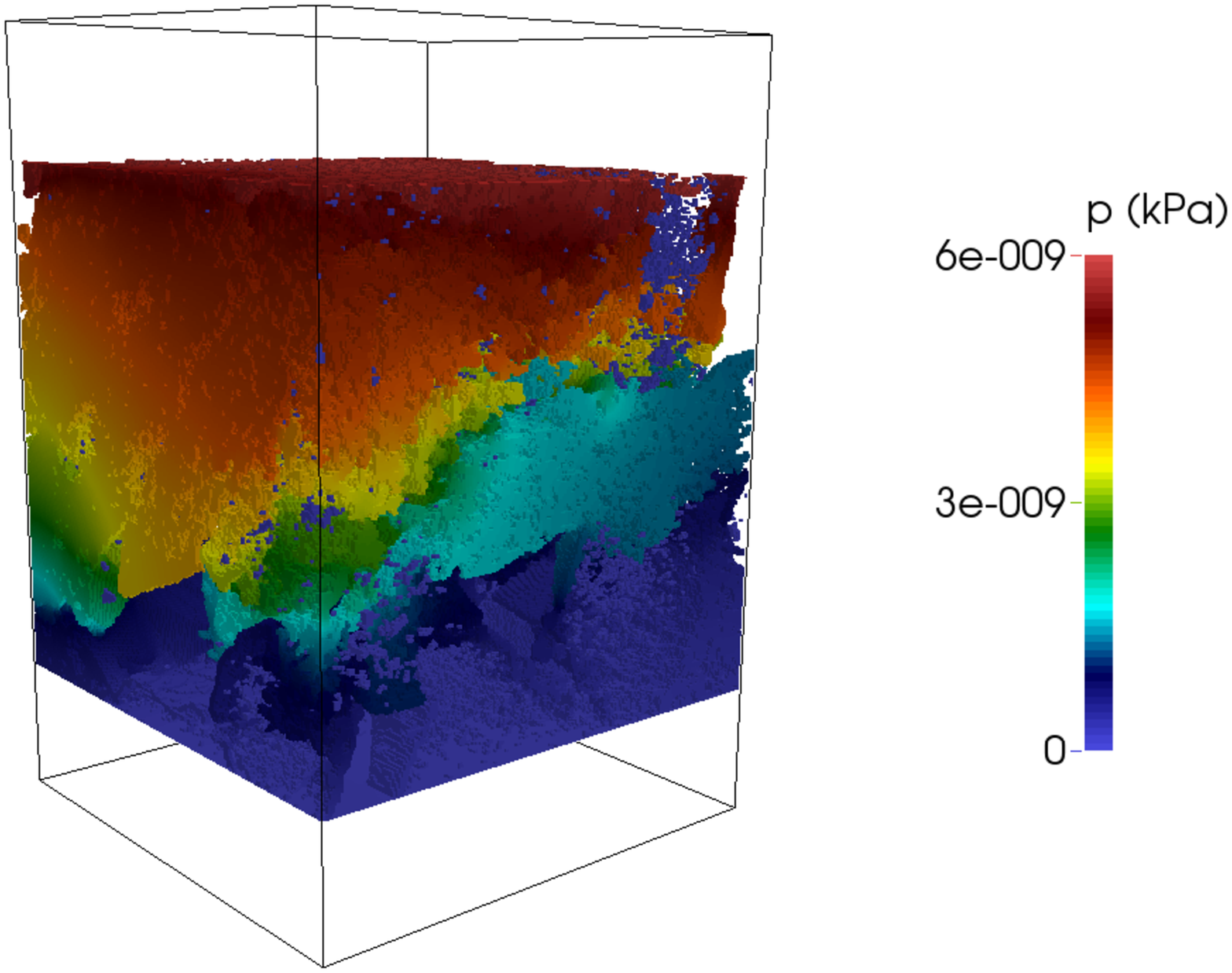}
\caption{Pressure, rock}
\label{fig:Rock_pressure_rescaled}
\end{subfigure}
\begin{subfigure}[c]{0.495\textwidth}
\includegraphics[width = \textwidth]{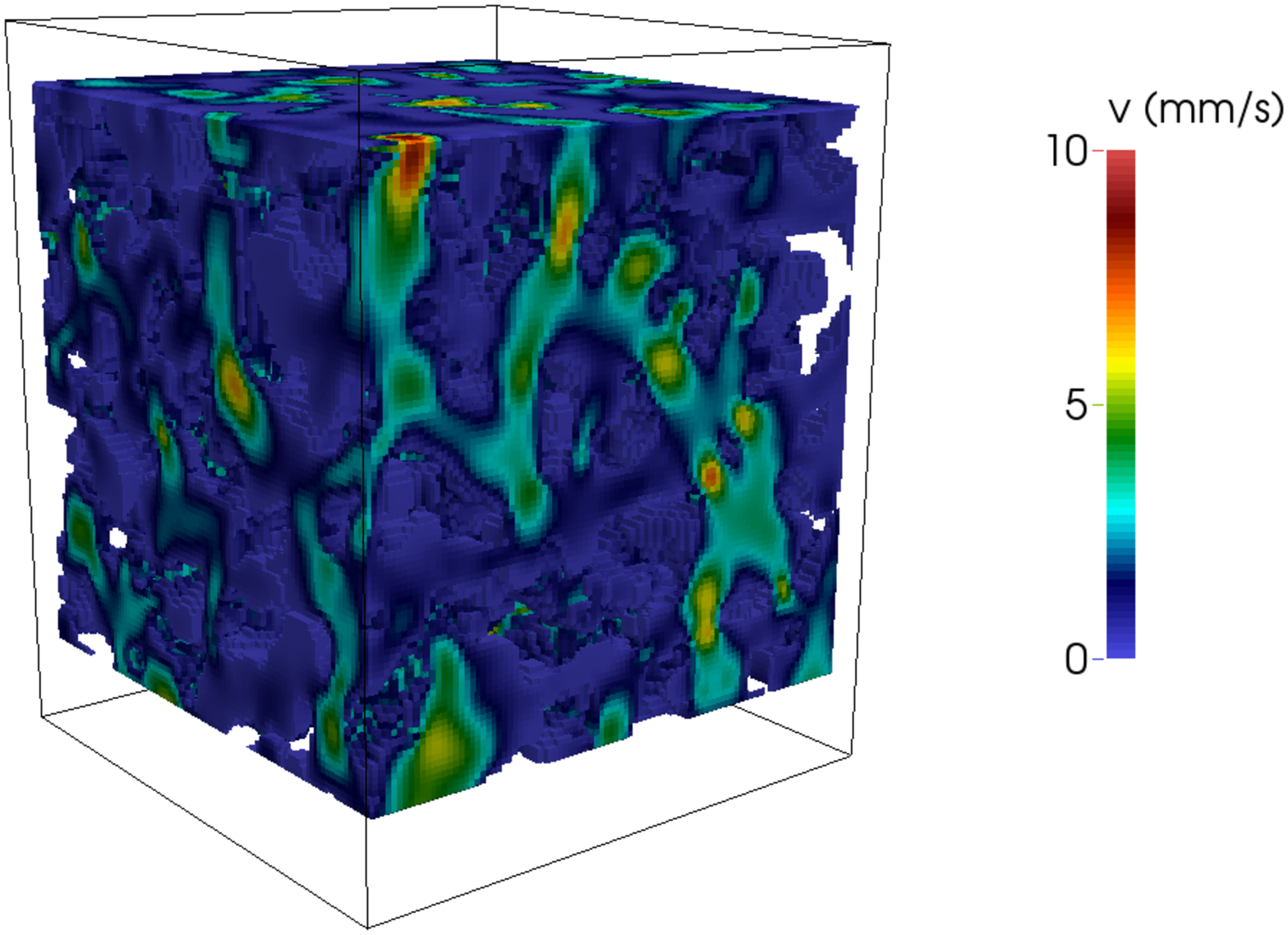}
\caption{Velocity magnitude, membrane}
\label{fig:Velocity_cut_perm_1_all}
\end{subfigure}
\begin{subfigure}[c]{0.495\textwidth}
\includegraphics[width = \textwidth]{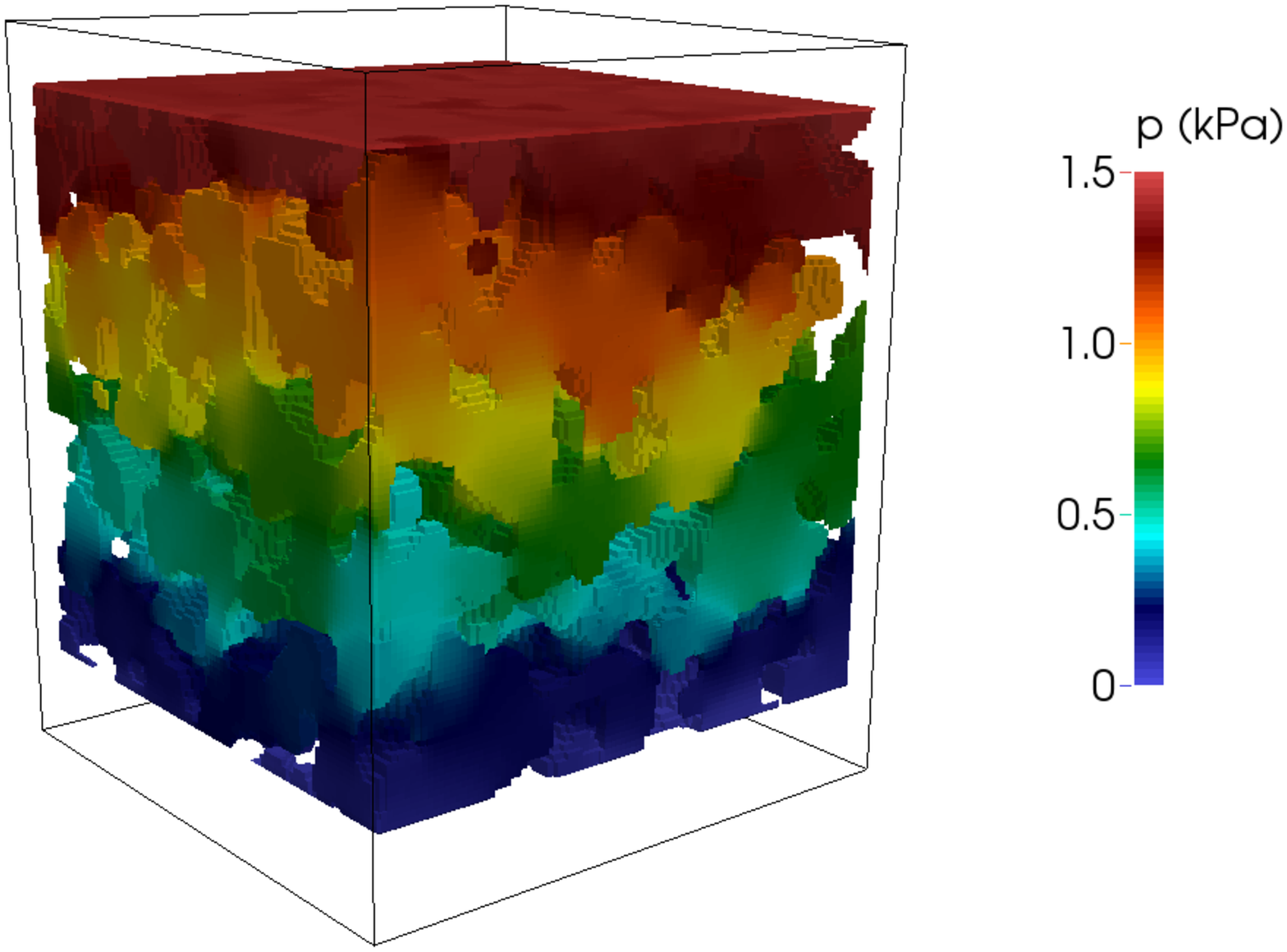}
\caption{Pressure, membrane}
\label{fig:Pressure_cut_perm_1_all}
\end{subfigure}
\caption{Velocity magnitude, \subref{fig:Rock_velocity_rescaled} and \subref{fig:Velocity_cut_perm_1_all},  and pressure, \subref{fig:Rock_pressure_rescaled} and \subref{fig:Pressure_cut_perm_1_all},  fields, measured in $\mathrm{mm}/\mathrm{s}$ and $\mathrm{kPa}$ respectively for both the computational domains considered.  Due to our assumption that the adsorption of the solute does not alter the geometry, these remain constant throughout the experimental time frame. The black boxes shows the outline for the entire computational domain, where we exclude the additional voxels embedded at the inlet and the outlet for 3D plotting purposes. }
\label{fig:Velocity_pressure}
\end{figure}

As described in Section~\ref{sec:numerical_methods}, we are first required to solve for the fluid flow.  
We assume that the fluid generates no slip as it passes over the membrane, and we set the slip length, $\beta$, to be zero so that, by  \eqref{eq:slip_BCs},  there is zero normal and tangential velocity at the fluid--solid interface, $\Gamma$.  We set inflow velocities  to be typical for the application under consideration, and use  $V_{\text{in}} = 1$~$\mathrm{mm}/\mathrm{s}$ for the membrane geometry and  $V_{\text{in}} = 1.5 \times 10^{-4}$~$\mathrm{mm}/\mathrm{s}$ for the rock geometry. The parameters chosen  for the inlet velocities, and the fluid density and viscosity, yield small Reynolds numbers; $\mathrm{Re} =  4.2 \times 10^{-7}$ in the case of the rock geometry and $\mathrm{Re} = 7.83 \times 10^{-3}$ for the membrane geometry. Therefore, the fluid flow in both computational domains is in a Stokes regime.    Solving \eqref{eq:Navier_Stokes_equations_eq1_ND} and  \eqref{eq:Navier_Stokes_equations_eq2_ND}, along with the boundary conditions \eqref{eq:NS_BC_ND_1}~--~\eqref{eq:Slip_BC_ND}, numerically until steady--state is achieved, yields the solutions as reproduced in Figure~\ref{fig:Velocity_pressure}. Due to our assumption that the maximal possible number of adsorbed particles  is sufficiently small to ignore the effects of geometry modification, these remain constant through time. 
Examination of Figure~\ref{fig:Velocity_pressure} reveals the dependence of the local velocity field on the morphology of the computational domains.

\subsection{Contaminant transport}
We now turn our attention to solving the reactive flow. 
Illustrative parameters are used, and we employ the Henry isotherm, given by Equation~\eqref{eq:Henry_Isotherm}, for the rock geometry, and the Langmuir isotherm, given by Equation~\eqref{eq:Langmuir_Isotherm},  for the membrane geometry.   We choose  $c_{\text{in}}$ arbitrarily to be $2 \times 10^4$~$\mathrm{number}/\mathrm{mm}^3$ and we  set the initial concentration of fluid contaminant to be equal to the inflow boundary condition, so that $\hat{c}_0(\hat{\mathbf{x}})=1$ for $\hat{\mathbf{x}} \in \hat{\Omega}_f$, while the quantity of adsorbed contaminant is initially assumed to be zero, $\hat{m}_0(\mathbf{x} )=0$.  
Furthermore, for the membrane  geometry,  we set the dimensionless maximal surface concentration of adsorbed contaminant, $\hat{m}_{\infty}$, to be $10^{-4}$, which equates to a dimensional value of $m_{\infty} = 0.014$~$\mathrm{number}/\mathrm{mm}^2$.  
Due to the form of the equations, the choice of $c_{\text{in}}$ does not influence the dimensionless system of equations describing the transport and reaction of the contaminant given by \eqref{eq:Convection_Diffusion_Equation_ND} along with the boundary conditions \eqref{eq:Reactive_BC_ND}~--~\eqref{eq:Frumkin_ND}, except for through the parameter $\hat{m}_{\infty}$. Therefore, increasing or decreasing $c_{\text{in}}$ for fixed $\hat{m}_{\infty}$ purely scales the dimensional  concentration (both fluid and adsorbed) by a constant factor.

\begin{figure}
\centering
\includegraphics[width = \textwidth]{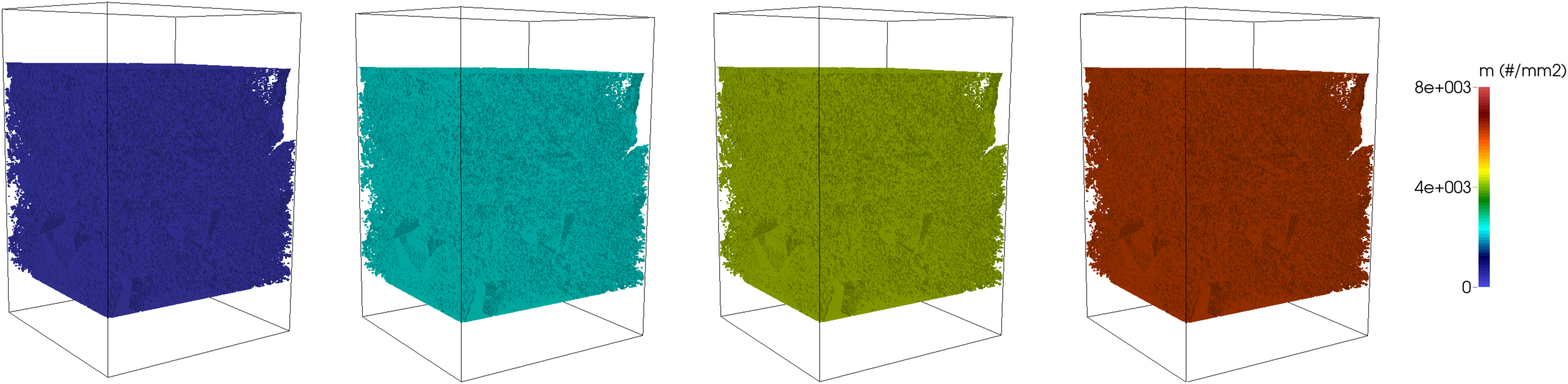}
\put(-420, -10){$t=0 $~$\mathrm{s}$}
\put(-350, -10){$t=1.2 \times 10^{3}$~$\mathrm{s}$}
\put(-235, -10){$t=2.4 \times 10^{3}$~$\mathrm{s}$}
\put(-125, -10){$t=3.8 \times 10^{3}$~$\mathrm{s}$}
\caption{The retained concentration of contaminant, $m$, for the rock geometry at 1200 second time intervals where $\mathrm{Da}_{a}  = 0.1$,  $\mathrm{Da}_{d}=   0.001$ and $\mathrm{Pe} = 2.0 \times 10^{-5}$. The fluid concentration remains effectively constant throughout the domain and so this is not plotted.}
\label{fig:Rock_geometry_retained} 
\end{figure}

\begin{figure}
\includegraphics[width = \textwidth]{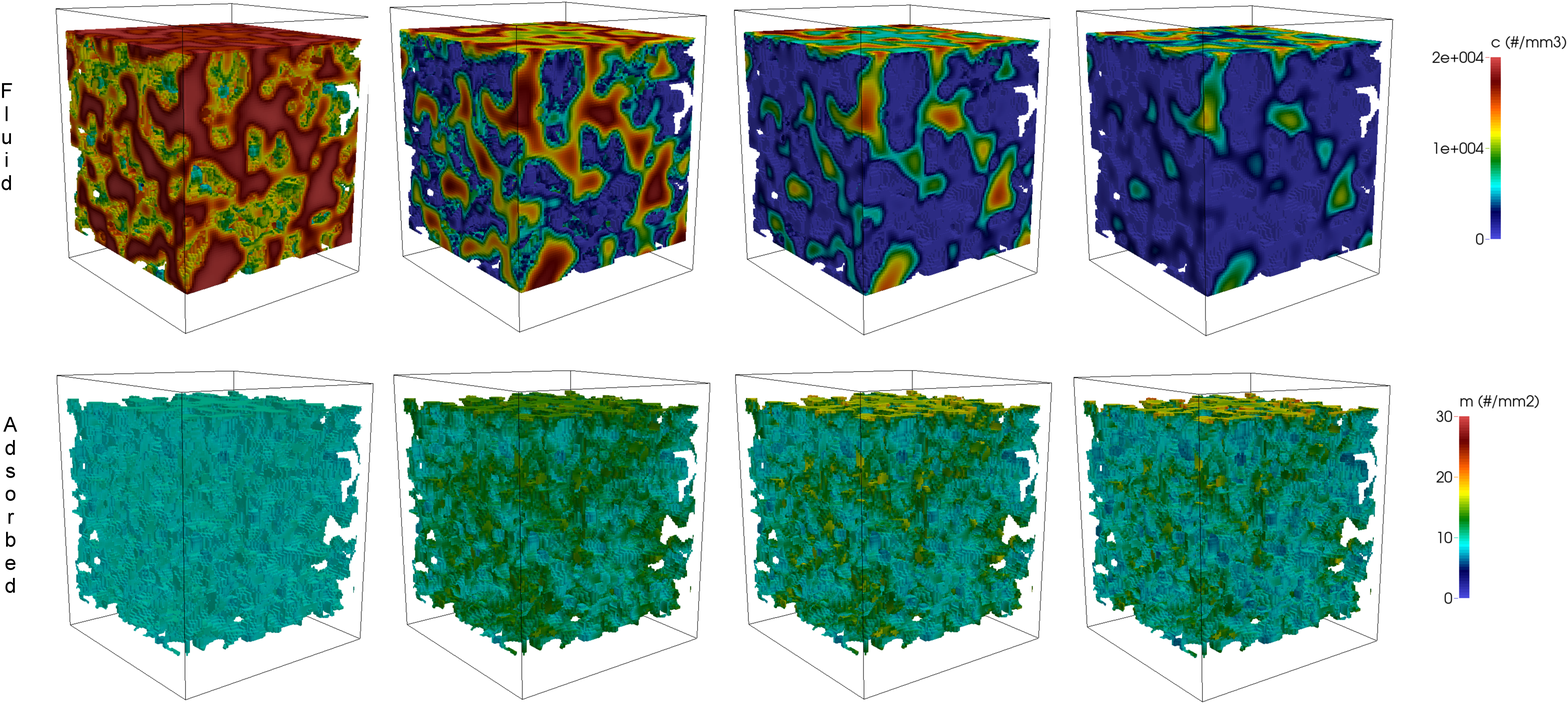}
\put(-425, -10){$t=5 \times 10^{-5}$~$\mathrm{s}$}
\put(-325, -10){$t=1 \times 10^{-4}$~$\mathrm{s}$}
\put(-235, -10){$t=1.5 \times 10^{-4}$~$\mathrm{s}$}
\put(-125, -10){$t=2 \times 10^{-4}$~$\mathrm{s}$}
\caption{Numerical results  at $5 \times 10^{-5}$~$\mathrm{s}$ intervals for the fluid concentration, $c$, and the adsorbed concentration, $m$, of contaminant in the membrane geometry. }
\label{fig:Simulation_1_conc_retained_conc}
\end{figure}

Figure~\ref{fig:Rock_geometry_retained} illustrates the concentration of the contaminant on the solid surface over time for the rock geometry,  where we use $\mathrm{Da}_{a}  = 0.1$,  $\mathrm{Da}_{d}=   0.001$ and $\mathrm{Pe} = 2.0 \times 10^{-5}$. Due to the slow rate of reaction and small P\'eclet number, the quantity of solute adsorbed over this time period is not large enough to significantly alter the fluid concentration over the time period under consideration.  Furthermore, as the rate of reaction is much smaller than the mass transport rate, the adsorbed concentration remains spatially homogeneous. 

In contrast, with  the parameters $\mathrm{Pe} = 10$ and $ \mathrm{Da}_a = \mathrm{Da}_d = 10$ for the membrane geometry, we see significant spatial heterogeneity in both the dissolved and adsorbed concentrations, as illustrated in Figure~\ref{fig:Simulation_1_conc_retained_conc}.  As time progresses, the fast rates of reaction result in a depletion of the dissolved  concentration at the pore wall and a  dependence of both the dissolved and adsorbed concentrations on the local membrane morphology.

\section{Conclusions}
\label{sec:Conclusion_Discussion}

We have presented an algorithm for solving    solute transport  at the pore--scale within a resolved porous medium, with reversible surface adsorption at the pore wall.   A pore--scale description of reactive transport, as opposed to a description at the Darcy scale,  allows for a very accurate representation of the processes of interest. 
The system of equations comprise the NS equations and  a CD equation, with Robin boundary conditions coupled to an ODE accounting for the surface reactions.  Assuming that  each particle is sufficiently small in size not to alter the flow of the fluid within the computational domain, and that its reaction at the wall does not significantly alter the pore--scale geometry, there is  a one--way coupling between the NS and CD systems of equations.   Although, for simplicity, we  consider one species of solute and  examine only surface reactions, extension to several different species of solute with both volumetric and surface reactions is straight forward and implemented within our software package Pore--Chem.  
 The algorithm presented employs a FV method,  and in this paper we have particularly focused on the discretization method used to solve the reactive  boundary conditions for the adsorption and desorption at the interface. 
 Illustrative numerical results, using our software package Pore--Chem, are presented on two separate geometries. The first of these  is a 3D $\mu$--CT image of a piece of Palatine Sandstone rock, while the second geometry is virtually generated within GeoDict \citep{Geodict} to be representative of a commercially available functionalized membrane.   The results demonstrate the potential of such a numerical package, with the ability to solve reactive transport directly on images and on virtually generated geometries, in further progressing the understanding of the interplay between the transport and reaction rates at the pore--scale.  In a future publication, currently in preparation, we will investigate the influence of the computational domain morphology on the reaction dynamics, and investigate the effect of different parameter regimes on the numerical results and quantities of interest. 

The advantages of using a pore--scale description are multiple. Firstly, it allows us to simulate reactive transport over a range of different parameter regimes, and in particular, outside the applicability region of the equivalent upscaled model. In  highly disordered media the P\'eclet number can significantly vary locally, which poses problems for asymptotic upscaling methods, but not for a pore--scale description.   Secondly, different kinetic models for the reactions can be used without the need to re--perform the upscaling procedure.    In the future we plan to extend the algorithm in order to solve coupled multiscale problems using the heterogeneous multiscale method, in a similar manner to   \citet{Battiato2011Hybrid} and \citet{Iliev2014Multiscale}.  Such a development will enable problems at larger spatial scales to be considered, which could aid further research into the influence of pore--scale processes in a number of highly interesting and important research applications.

\bibliography{AWR150707}
\end{document}